\documentclass{aa}  
\usepackage{longtable}
\usepackage{color}
\usepackage[dvipsnames]{xcolor}
\usepackage{hyperref}
\hypersetup{colorlinks=true, linkcolor=blue, urlcolor=blue, citecolor=blue}
\usepackage[all]{hypcap}
\usepackage{graphicx}
\usepackage[varg]{txfonts}
\usepackage{natbib}
\usepackage{enumitem}
\begin{document} 

   \title{The most massive stars in very young star clusters with a limited mass: Evidence favours significant self-regulation in the star formation processes}
    \titlerunning{The most massive stars in very young star clusters with a limited mass}
   \subtitle{}

  \author{Zhiqiang Yan\inst{1} \fnmsep \inst{2}
          \and
          Tereza Jerabkova\inst{3}
          \and
          Pavel Kroupa \inst{4} \fnmsep \inst{5}
          }

  \institute{
  School of Astronomy and Space Science, Nanjing University, Nanjing 210093, People's Republic of China
  \\ Emails: yan@nju.edu.cn; tereza.jerabkova@eso.org; pkroupa@uni-bonn.de
  \and 
  Key Laboratory of Modern Astronomy and Astrophysics (Nanjing University), Ministry of Education, Nanjing 210093, People’s Republic of China
  \and
  European Southern Observatory, Karl-Schwarzschild-Straße 2, 85748 Garching bei M{\"u}nchen, Germany
  \and
  Helmholtz-Institut f{\"u}r Strahlen- und Kernphysik (HISKP), Universität Bonn, Nussallee 14–16, 53115 Bonn, Germany
  \and
  Charles University in Prague, Faculty of Mathematics and Physics, Astronomical Institute, V Hole{\v s}ovi{\v c}k{\'a}ch 2, CZ-180 00 Praha 8, Czech Republic
  }
  
  \date{Received 8 September 2022 / Accepted 16 November 2022}

  \abstract
   {The stellar initial mass function (IMF) is commonly interpreted to be a scale-invariant probability density distribution function (PDF) such that many small clusters yield the same IMF as one massive cluster of the same combined number of stars. Observations of the galaxy-wide IMF challenge this as dwarf galaxies do not form as many massive stars as expected. This indicates a highly self-regulated star formation process in which stellar masses are not stochastically sampled from the IMF and are instead related to the environment of star formation. Here, the nature of star formation is studied using the relation between the most massive star born in a star cluster and its parental stellar cluster mass (the $m_{\rm max}$--$M_{\rm ecl}$ relation). This relation has been argued to be a statistical effect if stars are sampled randomly from the IMF. The tightness of the relation, given a set of young clusters compiled from the literature, probes this scenario. By comparing the observed $m_{\rm max}$--$M_{\rm ecl}$ distribution with synthetic star clusters with stochastically sampled stellar masses, we find that the expected dispersion of the mock observations is much larger than the observed dispersion. This conclusion depends on the accuracy of the mass uncertainties. Assuming that the uncertainties from the literature are correct, our test rejects the hypothesis that the IMF is a PDF at a more than $4.5\sigma$ confidence level. Alternatively, we provide a deterministic stellar mass sampling tool which reproduces the observed $m_{\rm max}$--$M_{\rm ecl}$ distribution and compares well with the luminosities of star-forming molecular clumps. In addition, we find that there is a significant flattening of the $m_{\rm max}$--$M_{\rm ecl}$ relation near $m_{\rm max}=13~M_\odot$. This may suggest strong feedback of stars more massive than about $13~M_\odot$ and/or the ejections of the most massive stars from young clusters in the mass range 63 to $400~M_\odot$ to be likely important physical processes in forming clusters.
   }


   \maketitle

\section{Introduction}\label{sec:intro}

The nature of the discreteness of the stellar initial mass distribution function (IMF) is important to consider. Different methods to sample stellar masses from an IMF have been discussed and applied in an increasing number of recent studies of stellar systems with a low mass or a low star formation rate (SFR) and with a high resolution of time or mass, starting to account for the evolution of individual stars (e.g. \citealt{2017MNRAS.466.1903G,2017MNRAS.466..407S,2017MNRAS.471.2151H,2019MNRAS.482.1304E,2019MNRAS.483.3363H,2020MNRAS.492....8A,2020ApJ...891....2L,2021PASJ...73.1036H,2021MNRAS.501.5597G,2021MNRAS.506.3882S,2022MNRAS.509.5938H}).

It has been long known that simple stellar population (SSP) modelling codes, for example, GALEV \citep{2003A&A...401.1063A} and Starburst99 \citep{2005ApJ...621..695V} neglect the discrete nature of the stellar mass distribution. This introduces a systematic offset in the colour prediction for star clusters that are young ($<80~{\rm Myr}$) or have a low mass ($<10^6~M_\odot$, where $M_\odot$ is the solar mass, \citealt{2004A&A...413..145C}). 
Ignoring the discreteness of the IMF significantly affects the inferred properties of low-mass systems, such as the evolution of a star cluster \citep{2021A&A...655A..71W,2022A&A...660A..61D}, the age and mass determinations of star clusters, introducing errors in cluster ages as high as an order of magnitude \citep{2009A&A...507L...5P}.

In the dynamical simulations of galaxies, an ensemble of stars called a ``star particle'' with a mass resolution of, for example, $10^5~M_\odot$, is often tracked instead of individual stars. In the simulation of single dwarf galaxies (and in recent years, the cosmological galaxy simulations that are pushing to higher and higher resolutions to study ultra-faint galaxies), the number of stars in a galaxy or a ``star particle'' becomes low enough that the IMF can no longer be considered as a continuous function. A fraction of a supernova explosion in a time step obtains when the integrated number of stars between two mass limits of the IMF (given by the considered time step and the stellar mass--lifetime relation, cf. \citealt{2019A&A...629A..93Y} their Fig.~3) is less than one \citep{2016A&A...588A..21R,2020MNRAS.492....8A}. It has been demonstrated that a more realistic description of a stellar population with a list of sampled stellar masses alters the property and the evolution of the simulated galaxy compared to a model with a continuous energy injection due to the smooth IMF assumption, affecting the galactic SFR \citep{2018MNRAS.480.1666S}, the strength of photoionization \citep{2021MNRAS.502.5417S}, and chemical composition \citep{2008MNRAS.390..582C}. 

However, an important question that has not been resolved is how to sample stellar masses from a given IMF.
There are two extreme sampling methods – stochastic sampling (a.k.a. random sampling) and optimal sampling. Stochastic sampling is based on the hypothesis that the formation of every single star is purely an independent event and the mass of the star is randomly sampled from a probability density distribution function (PDF). That is, the IMF is considered to be a PDF.
On the other end, optimal sampling \citep{2013pss5.book..115K,2015A&A...582A..93S} assumes that the formation of all the stars (or at least all the massive stars that matter observationally) in a star cluster is strongly correlated due to self-regulation of the star formation process. Thus, a deterministic relationship exists between the mass of a star cluster and the mass of every single (massive) star within that cluster.

Theoretically, the sampling method relates to the fundamental astrophysics knowledge of the formation process of massive stars that is not well understood.
If the star formation process is sensitive to the initial conditions of the molecular cloud, for example, angular momentum, shock wave behaviour, etc, then stochastic sampling will be a reasonable approximation. 
For example, in the competitive accretion model \citep{1997MNRAS.285..201B,1998MNRAS.298...93B,2001MNRAS.324..573B}, the stellar mass is gathered during the star-formation process as the parent cloud is randomly influenced by dynamical interactions. 
On the other hand, it is also possible that massive star formation is strongly regulated by feedback processes such as winds, outflows, radiation, accretion shocks, and the magnetic field of the parent cloud \citep{1987ARA&A..25...23S,1995ApJ...438L..41S,1996ApJ...464..256A,2006ApJ...640L.187L,2007IAUS..237..141C} such that the random initial conditions are not as important.
Then the stars formed in a single star formation event may follow the IMF tightly with much more certainty than the random scenario.
Another possibility is that a significant self-regulated formation of star clusters does not even require a feedback regulation as nature can automatically form the power-law distribution through scale-invariance, self-similarity, and iterative processes. For example, in preferential attachment phenomena and diffusion-limited aggregation, a large number of random processes (instead of a single stochastical event) determines the outcome which becomes highly deterministic. The state of art star cluster formation simulation, STARFORGE, appears to support this view \citep{2022arXiv220510413G}. Despite the chaotic and complex physical processes involved, their simulated maximum stellar masses in embedded star clusters\footnote{An embedded cluster refers to a correlated star formation event, i.e., a gravitationally driven collective process of transformation of the interstellar gaseous matter into stars in molecular-cloud overdensities on a spatial scale of about one pc and within about one Myr \citep{2003ARA&A..41...57L,2013pss5.book..115K,2016AJ....151....5M}. The value $M_{\mathrm{ecl}}$ is the mass of all stars formed in the embedded cluster.} seem to have too little scatter to be explained by stochastic sampling schemes with strong evidence of a correlation between the embedded star cluster mass and the mass of the most massive star ($m_{\rm max}$--$M_{\mathrm{ecl}}$ relation), being consistent with optimal sampling.

A variety of observational clues suggesting a highly self-regulated star formation process instead of a statistical process has been discussed in \citet[table 1 therein]{2015CaJPh..93..169K}.
A non-trivial correlation between $m_{\rm max}$ and $M_{\mathrm{ecl}}$ is evident through empirical \citep{1982MNRAS.200..159L,2003ASPC..287...65L}, analytical \citep{2000ApJ...539..342E}, and numerical \citep{2003MNRAS.343..413B,2004MNRAS.349..735B} studies.
The tight relation is particularly emphasised by \citet{2006MNRAS.365.1333W,2010MNRAS.401..275W,2013MNRAS.434...84W} and reinforced by \citet{2017ApJ...834...94S} and \citet{2017A&A...607A.126Y}, suggesting a significant self-regulation in the star formation process.
\cite{2016A&A...592A..54A} find that stars form in molecular-cloud overdensities from inflows along filaments with remarkably similar cross sections of about 0.1~pc radii, suggesting that star formation may be governed by relatively simple universal laws. A direct comparison of the fibre networks in B213 in Taurus and NGC 1333 region in Perseus suggests that the differences between the classically distinguished isolated and clustered star-formation scenarios might originate from the same type of fibre with different densities. And the interaction between fibres may be responsible for the
formation of massive cores and stars \citep{2017A&A...606A.123H}.
The recent observations also reveal that there are regions with numerous low-mass correlated star formation events (CSFEs, aka embedded clusters) that do not possess any massive star. For example, the whole southern part of the Orion southern cloud, L1641, has not formed a single massive star, despite thousands of stars having formed there (see e.g., \citealt{2012ApJ...752...59H,2013ApJ...764..114H} and \citealt{2016AJ....151....5M}). 
This is consistent with the $m_{\rm max}$--$M_{\mathrm{ecl}}$ relation because the southern cloud has been forming low-mass embedded clusters only.
These CSFEs have radii consistent with the filament cross sections \citep{2012A&A...543A...8M}.
In addition, \citet{2012ApJ...761..124G} and \citet{2013MNRAS.435.2604P} study the cluster mass distribution at different galactic radii of the flocculent galaxy M 33 and conclude that their mass is not randomly distributed.

Motivated by the above observations, \citet[their section 2.2]{2013pss5.book..115K} suggest that the optimal sampling hypothesis better describes nature.
In particular, optimal sampling is crucial to understand the top-light IMF (lack of massive stars compared to canonical IMF) of dwarf galaxies, evident by their H$\alpha$-to-FUV luminosity ratio \citep{2007ApJ...671.1550P,2009ApJ...706..599L,2009MNRAS.395..394P},
the mass-metallicity relation of galaxies \citep{2007MNRAS.375..673K},
and their $\alpha$-element enhancement in comparison with the Milky Way (MW, \citealt{2009A&A...499..711R,2020A&A...637A..68Y,2021ApJ...910..114M,2021NatAs...5.1247M}). With optimal sampling, the dwarf galaxies with a low SFR do not form massive stars \citep[their figure 7]{2017A&A...607A.126Y}. This is not the case in stochastic sampling where the formation of massive stars happens randomly throughout the formation history of a galaxy such that the ratio between low-mass and massive stars ever formed (and therefore the chemical evolution) does not depend on the mass of a galaxy. 

Until now, the more common way of generating stellar masses for a star system in astrophysical simulations has been the stochastic generation \citep{2006ApJ...648..572E,2012ApJ...745..145D}.
After all, the observed star clusters with similar masses do not appear to repeat themselves but appear to have random differences in their stellar mass distributions. The problem is whether these differences are intrinsically due to the random nature of star formation or if they come from 
hidden parameters such as the metallicity \citep{2012MNRAS.422.2246M} and the angular momentum of the parental molecular cloud, through the ejection of massive stars out of their birth clusters \citep{2015ApJ...805...92O,2018MNRAS.481..153O}. The observation that most massive stars have an unresolved companion star with, apparently, a random mass ratio of the two stars \citep{2017ApJS..230...15M} also introduces the complication of binary stellar evolution.
Furthermore, the total stellar mass of an embedded star cluster is uncertain due to extinction and other measurement uncertainties. 
Therefore, apparent randomness does not necessarily suggest a random star formation process and a more detailed investigation is required.

The present work provides a new and IMF independent statistical test on the degree of self-regulation in the star formation process. 
The observational dataset for the $m_{\rm max}$--$M_{\mathrm{ecl}}$ relation is given in Section~\ref{sec: data}.
Our statistical test method and IMF assumptions are described in Section~\ref{sec: Method}.
Statistical tests for random and optimal sampling methods are performed in Section~\ref{sec: Test stochastic sampling} and \ref{sec: Test optimal sampling}, respectively.
In addition, we find a large $M_{\mathrm{ecl}}$ variance around $m_{\rm max}=13~M_\odot$ for the observed star clusters. This is discussed in Section~\ref{sec: Scatter of the data}. 
We further discuss the uncertainties of this study in Section~\ref{sec: Uncertainty analyses}, discuss previous studies and the implementation of different sampling methods in Section~\ref{sec: Discussion}, and give conclusions in Section~\ref{sec: Conclusion}.

\section{Data selection}\label{sec: data}

The observed values of ``embedded'' star cluster masses, $M_{\mathrm{ecl}}$, and the mass of the most massive star in that star cluster, $m_{\rm max}$, for star clusters with an age smaller than 5 Myr are considered. Here we apply a compilation of data from \cite{2011ApJ...727...64K}, \citet[hereafter WKP13]{2013MNRAS.434...84W}, and \cite{2017ApJ...834...94S}. The dataset is listed in Table~\ref{table: Literature data} and details are provided in Section~\ref{sec: Mass uncertainties}.

Compared to the dataset in WKP13 and \citet{2017A&A...607A.126Y}, we apply an additional data selection on the age of a star cluster and the number of stars in a cluster.
The age limit reduces the possibility of having an old cluster in the sample that is highly evolved dynamically or with the most massive star already dead in a supernova explosion. The most massive stars will disappear within the first few Myr of cluster evolution through dynamical ejections \citep{2016A&A...590A.107O} and stellar evolution, while massive stars also merge \citep{2018MNRAS.481..153O}. The shortest lifetime of a star is no less than about two to three Myr (\citealt{2013MNRAS.433.1114Y}, \citealt[their figure 3]{2019A&A...629A..93Y}, \citealt[their figure 6]{2022MNRAS.516.4052H}). Although WKP13 culled data only with the centre of age estimation smaller than 5 Myr, some of the age estimations have a large uncertainty and may have an age as old as 9 Myr within the standard deviation of the age estimation. To reduce uncertainties introduced by older star clusters, we exclude the tail of the age distribution from the above data reservoir, excluding all the star clusters with an age estimation upper limit above 5 Myr. The young ages of the clusters ascertain their expansion through gas expulsion is minimised (to 5 pc for a velocity dispersion of 1 km/s) such that the loss of stars is small \citep{2001MNRAS.321..699K,2003MNRAS.346..343K}. Ejections and mergers of massive stars are, however, likely \citep{2012MNRAS.424...65O,2018MNRAS.481..153O,2018A&A...612A..74K}.
In addition, clusters with less than 20 detected members (or if no information is provided) are excluded. Clusters with fewer members are strongly affected by dynamical evaporation (cf. \citealt[their section 2.1]{2003ARA&A..41...57L}) and can bias our sample (see Section~\ref{sec: IMF bias}). In addition, the $m_{\rm max}$--$M_{\mathrm{ecl}}$ relation for low-mass clusters is not sensitive enough to test the sampling method given the large mass measurement uncertainties. For massive clusters, a low number of single star detections indicate different mass estimation methods (e.g. use dynamical cluster mass instead of summing up the masses of individual stars, see Section~\ref{sec: IMF bias}) and, thus, different mass estimation uncertainties compared to most star clusters. Therefore, we uniformly exclude such clusters from our dataset.

After the selection, there are in total 100 clusters left. A list of them is provided in Table~\ref{table: Literature data} and plotted in Fig.~\ref{fig: RSMmaxMecl}.
The average observational uncertainties in the logarithmic scale for $M_{\rm ecl}$ and $m_{\rm max}$ are $+0.289$, $-0.313$ dex and $+0.143$, $-0.180$ dex, respectively. The accuracy of these uncertainties is discussed in Section~\ref{sec: Mass uncertainties}.
Scaling relations are applied to calculate the pre-cluster molecular cloud mass and core density as well as the bolometric luminosity of the most massive star, which can be more easily compared with observations (e.g. \citealt{2022MNRAS.510.3389U}). The results
are shown in Appendix~\ref{Appendix: Properties of the star forming clouds}.

The errors and biases of the compiled dataset affect our results as discussed in Section~\ref{sec: Uncertainty analyses}. 
In addition, we acknowledge that the stellar mass upper limit is highly uncertain. There are studies suggesting stars of mass $>300~M_\odot$ \citep{2010MNRAS.408..731C,2018Sci...359...69S,2022MNRAS.516.4052H} that might be a result of mergers of massive stars \citep{2012MNRAS.426.1416B,2018MNRAS.481..153O}. However, the mass estimation of massive stars is not the topic of the present study and our conclusion is not affected if we adopt a higher stellar mass estimation.
For the present tests, we apply the compiled data along with the measurement errors given by the literature sources. A homogeneous set of observations with higher completeness (the current dataset includes less than about $2\%$ of all the star clusters in the MW, c.f. \citealt{2010MNRAS.401..275W}) and a more reliable mass uncertainty analysis will, in the future, greatly improve the robustness of our study.

\section{Test method}\label{sec: Method}

\subsection{Statistical test}\label{sec: Statistical test}

As has been pointed out in WKP13, the scatter of the observed $m_{\rm max}$--$M_{\mathrm{ecl}}$ relation appears to be consistent with no intrinsic scatter, which suggests a high degree of self-regulation. 
To test if nature is indeed significantly more regulated than the stochastic sampling hypothesis, we compare the scatter of the observed with the synthesised $m_{\rm max}$--$M_{\mathrm{ecl}}$ relation generated by stochastic sampling.

With forward modelling, a large number of star clusters and their mock observational errors are stochastically generated to form the expected $m_{\rm max}$--$M_{\mathrm{ecl}}$ density distribution.
The mock clusters are set to have a cluster mass distribution similar to the observational dataset, which has a relatively uniform distribution of mass on the logarithmic scale. Therefore, we generate $10^6$ star clusters containing $N$ stars, where $N$ is randomly selected from 20 to $10^6$ with a uniform distribution on the logarithmic scale (we note that the star number and mass distribution is similar, see WKP13 their section~2.2). The observed and synthesised cluster mass distribution is different from the embedded cluster mass distribution in nature with a power-index of -2 \citep{2003ARA&A..41...57L} due to an observational bias towards more massive star clusters. 

Given the number of stars in each star cluster, the masses of stars are sampled stochastically (e.g. \citealt{2013pss5.book..115K}) from a PDF, that is, the assumed IMF.
The model results depend on both the sampling method and the IMF. 
To test the sampling method in a way that the conclusion does not depend on the assumed shape of the IMF, the present work applies two free parameters in the IMF, the high-mass IMF slope, $\alpha_3$, and the stellar mass upper limit, $m_{\rm up}$. We search for the two parameters that best reproduce the observed $m_{\rm max}$--$M_{\mathrm{ecl}}$ relation (Section~\ref{sec: IMF}) and compare the scatter of the mock star clusters and the data on the $m_{\rm max}$--$M_{\mathrm{ecl}}$ plot for the first time.
A different IMF prescription, such as a lognormal IMF for low-mass stars \citep{2003PASP..115..763C}, would introduce a systematic change on the $m_{\rm max}$--$M_{\mathrm{ecl}}$ relation similar to the effect of varying the IMF parameters ($\alpha_3$ and $m_{\rm up}$) and would not affect our conclusions.

To compare the synthetic star clusters with observations, mock measurement errors of $m_{\rm max}$ and $M_{\mathrm{ecl}}$ are added to the sampled clusters for the first time.
The two mass errors are correlated (see discussion in Section~\ref{sec: Mass uncertainties}) but, as a first step, we assume that they are independent and that they follow one-sided normal distributions with standard deviations being the average observational uncertainties from our complied dataset given in Section~\ref{sec: data}. 
Doing so results in rare cases with $m_{\rm max}>M_{\mathrm{ecl}}$ in mock observations.
Follow-up studies with a uniformly analysed observational dataset allowing accurate forward modelling of the errors are required to resolve this issue. 

With the observation and mock observation of synthetic star clusters directly comparable, a two-sample test can be performed on the 2-dimensional $m_{\rm max}$--$M_{\mathrm{ecl}}$ relation.
To do this, we first measure the variance of $m_{\rm max}$ values for each $M_{\mathrm{ecl}}$ bin for the simulated star clusters. Since there is no limit on the number of mock star clusters, the bin size can be arbitrarily small.
An observed star cluster with a known $M_{\mathrm{ecl}}$ can then be compared with the mean and the variance of all the mock star clusters in the mass bin that includes the value $M_{\mathrm{ecl}}$.
After doing this for all the observed star clusters, we can calculate the likelihood of observing a dataset as extreme as the one we compile from the literature if the assumed sampling method is true. Finally, the confidence level of rejecting the sampling hypothesis is given.

\subsection{IMF models}\label{sec: IMF}

Different IMF assumptions are considered to avoid the influence of the IMF model when testing the validity of the sampling methods.
The multipart power law IMF can be written in general as:
\begin{equation}\label{eq:xi}
\xi(m) =
\begin{cases} 
0, & m<m_{\rm low}, \\
k_1 m^{-\alpha_1}, & m_{\rm low} \leqslant m<m_{\rm turn}, \\ 
k_2 m^{-\alpha_2}, & m_{\rm turn} \leqslant m<m_{\rm turn_2}, \\
k_3 m^{-\alpha_3}, & m_{\rm turn_2} \leqslant m<m_{\rm up},\\
0, & m_{\rm up} \leqslant m.
\end{cases}
\end{equation}
The canonical \cite{2001MNRAS.322..231K} IMF is a two-part power law with $m_{\rm low},~m_{\rm turn},~m_{\rm turn_2} = 0.08,~0.5,~1~[M_\odot$], respectively, $\alpha_1=1.3$, $\alpha_2=\alpha_3=2.3$, and $m_{\rm up}=150~M_\odot$. Possible variations of $\alpha_1$ and $\alpha_2$ are discussed in Section~\ref{sec: IMF bias}. The normalisation parameters, $k_1$, $k_2$, and $k_3$ are adjusted such that the IMF is a continuous function.

The slope of the high mass IMF, $\alpha_3$, and the physical stellar mass upper limit, $m_{\rm up}$ ($=m_{\rm max*}$ in \citealt{2006MNRAS.365.1333W,2010MNRAS.401..275W}), affect the $m_{\rm max}$--$M_{\mathrm{ecl}}$ relation significantly and both have large uncertainties.
Different choices of these parameters are listed in Table~\ref{table: model names}. We note that $\alpha_3=2.68$ is similar to the Solar-neighbourhood IMF measured in \citet{1986FCPh...11....1S} and \citet{1993MNRAS.262..545K}. This is steeper than the IMF in star clusters (cf. \citealt{2010ARA&A..48..339B}), which is expected for a galaxy-wide IMF made by many star clusters \citep{2003ApJ...598.1076K}. The high-mass IMF slope for the field stars is not accurate due to a large uncertainty of the MW star formation history. A recent study shows that the Solar-neighbourhood IMF for massive stars can be consistent with the IMF in star clusters \citep{2019A&A...624L...1M}.
For initial stellar populations in star clusters, a flatter slope is favoured and $\alpha_3=2.3\pm0.7$
as is shown by \citet{2001MNRAS.322..231K,2002Sci...295...82K} and \citet{2010ARA&A..48..339B}. On the other hand, the estimation of $m_{\rm up}$ depends on the stellar-modelling of massive stars and is still under debate (see Section~\ref{sec:very massive stars}). We vary $m_{\rm up}$ to find the best fit with the data without a prior probability distribution.
\begin{table}
\caption{The best-fit IMF models}              
\label{table: model names}      
\centering                                      
\begin{tabular}{c c c}          
\hline\hline                        
Model name & $\alpha_3$ & $m_{\rm up}$ \\    
\hline                                   
   a23m150 & 2.3 & 150  \\      
   a23m100 & 2.3 & 100  \\
   a268m130 & 2.68 & 130  \\
\hline                                             
\end{tabular}
\tablefoot{The fiducial a23m150 model adopts the canonical IMF parameters. a23m100 is the best fit IMF model working with stochastic sampling with $m_{\rm up}$ being a free parameter. a268m130 is the best fit IMF model working with stochastic sampling with both $\alpha_3$ and $m_{\rm up}$ being free parameters. We note that the best-fit $m_{\rm up}$ value listed here is only a reflection of adopted literature mass estimations while higher mass estimations of extremely massive stars have been suggested \citep{2010MNRAS.408..731C,2018Sci...359...69S,2022MNRAS.516.4052H} but may be mergers \citep{2012MNRAS.426.1416B,2018MNRAS.481..153O}.}
\end{table}

\section{Tests of stochastic sampling}\label{sec: Test stochastic sampling}

\subsection{The mock $m_{\rm max}$--$M_{\mathrm{ecl}}$ relation from stochastic sampling}\label{sec: mock stochastic sampling results}

To discuss the degree of self-regulation in the star formation processes independent of the IMF shape, we not only test with a fiducial IMF model (a23m150 model in Table~\ref{table: model names}) but also allow freely variable IMF parameters and test parameter values that can best reproduce the observed $m_{\rm max}$--$M_{\mathrm{ecl}}$ relation (a23m100 and a268m130 models in Table~\ref{table: model names} shown in Section~\ref{sec:Model a23m100} and \ref{sec:Model a268m130}).

\begin{figure}[!hbt]
    \centering
    \includegraphics[width=9cm]{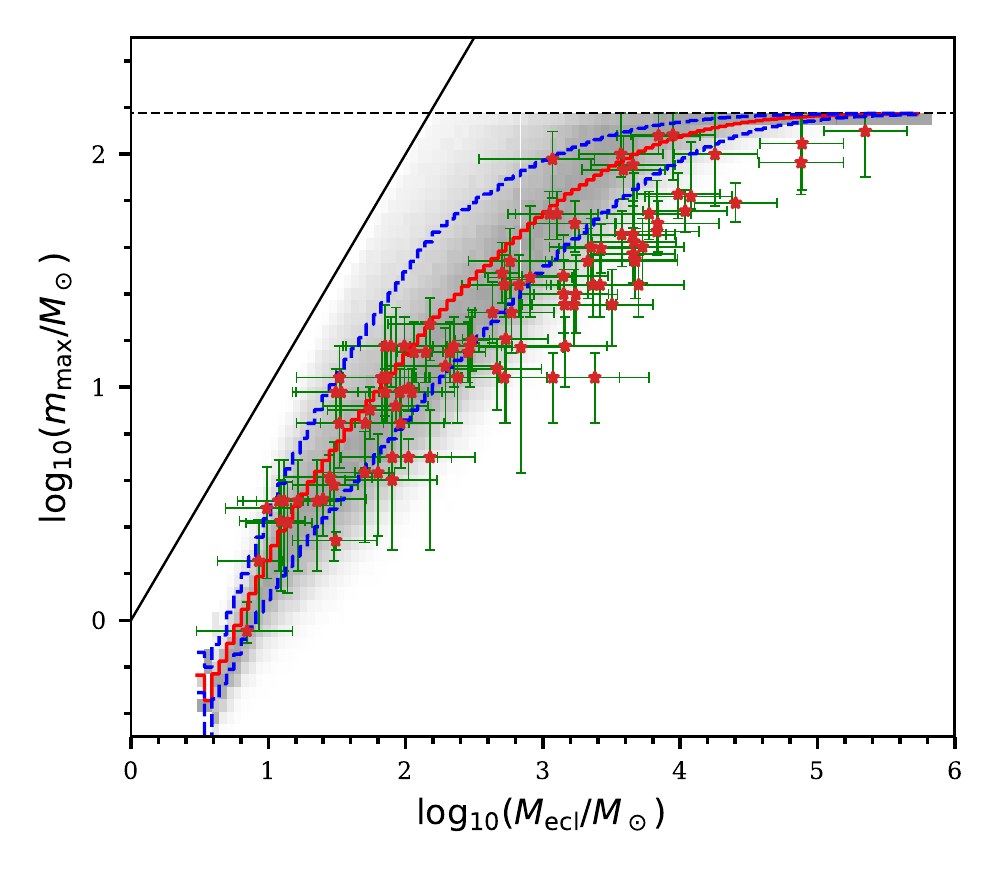}
    \caption{Most massive stellar mass to embedded cluster mass ($m_{\rm max}$--$M_{\mathrm{ecl}}$) relation. Red stars denote the observational data listed in Table~\ref{table: Literature data}. $10^6$ synthetic clusters are stochastically sampled from the canonical IMF (the a23m150 IMF model in Table~\ref{table: model names}), resulting in the density map. The darkness of the region is proportional to the number of the synthetic clusters in that region divided by the total number of synthetic clusters in the same $M_{\mathrm{ecl}}$ bin. The average $m_{\rm max}$ and standard deviation of $m_{\rm max}$ for each $M_{\rm ecl}$ bin are plotted as the red solid and the blue dashed curves, respectively. The solid line indicates the $M_{\mathrm{ecl}}=m_{\mathrm{str,max}}$ limit and the horizontal dashed line indicates the $m_{\rm up}$ limit of $150~M_{\odot}$ (see Section~\ref{sec:very massive stars}).}
    \label{fig: RSMmaxMecl}
\end{figure}

Fig.~\ref{fig: RSMmaxMecl} shows the distribution on the $m_{\rm max}$--$M_{\mathrm{ecl}}$ relation for $10^6$ randomly-sampled embedded clusters applying our fiducial IMF model parameters. The darkness of each region is proportional to the number of clusters in a $m_{\rm max}$--$M_{\mathrm{ecl}}$ bin divided by the total number of clusters in the same $M_{\mathrm{ecl}}$ bin. That is, the darkness denotes the distribution of $m_{\rm max}$ for a given $M_{\mathrm{ecl}}$. The mean and the standard deviation of $m_{\rm max}$ for each $M_{\rm ecl}$ bin are also given with the red solid and the blue dashed curves, respectively. Here, the single-sided standard deviations are calculated and shown, assuming that data above and below the mean $m_{\rm max}$ have different variances, but the upper and lower variances are nearly identical.

The observed star clusters more massive than about $100~M_\odot$ are systematically below the expected distribution as is pointed out in \citet[their section~3.2]{2010MNRAS.401..275W}. A top-light IMF (model a268m130) can result in a better agreement which we discuss in Section~\ref{sec: P-value test - stochastic sampling}. 
The mass uncertainties of the observations shown in Fig~\ref{fig: RSMmaxMecl} are large while the distribution of the measurement values is tight, indicating a small intrinsic uncertainty of the star formation process (disfavouring stochastic sampling), an overestimation of the mass uncertainty of observed clusters, and/or a selection bias of the observational dataset.

\begin{figure}[!hbt]
    \centering
    \includegraphics[width=9cm]{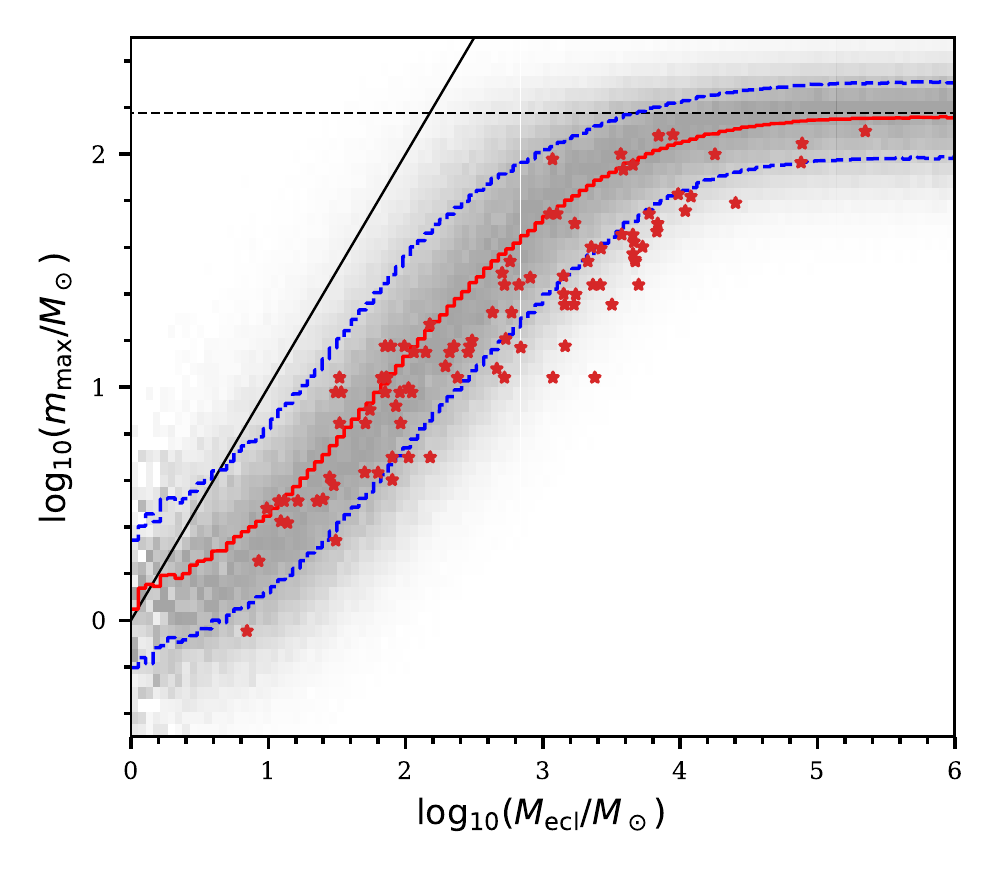}
    \caption{Same as Fig.~\ref{fig: RSMmaxMecl} with the same set of synthetic star clusters (model a23m150) but added with mock observational uncertainties.}
    \label{fig: RSMmaxMecl2}
\end{figure}

Fig.~\ref{fig: RSMmaxMecl2} shows the result after a random observational error is added to each synthetic star cluster.
The density distribution shows a larger scatter than the data. The amplitude of this scatter is determined by the sampling method and observational uncertainty. Changing the IMF parameters does not have a significant influence on the scatter but only influences the position and the orientation of the density distribution which is shown in the following section. 
We note that there are mock clusters with $M_{\mathrm{ecl}}<m_{\mathrm{str,max}}$ (points above the solid line in Fig.~\ref{fig: RSMmaxMecl2}) only because we assume for simplicity that the observational uncertainty of $M_{\mathrm{ecl}}$ and $m_{\mathrm{str,max}}$ are independent, which is not accurate, especially for low-mass clusters. However, given the statistical test method described in the following section, the error for the low-mass mock clusters affects our conclusions insignificantly as there are only three clusters in the dataset with $M_{\mathrm{ecl}}<10~M_\odot$.

\subsection{The p-value}\label{sec: P-value test - stochastic sampling}

Assuming the null hypothesis that nature behaves exactly like what stochastic sampling describes and the IMF being applied is correct, here we calculate the likelihood of observing a dataset as extreme as the one we compiled from the literature (i.e. the 100 star clusters listed in Table~\ref{table: Literature data}). We choose a test statistic, $n$, to be the number of data points in the region above the upper blue dashed curve (hereafter the upper region) or the number of points below the lower blue dashed curve (hereafter the lower region). The expected value of $n$ is $100\times0.159=15.9$ points, where 0.159 is the single-sided probability above the standard deviation.

If there are $n\leq15$ points in the upper or lower region, we calculate the likelihood of having $\leq n$ points in that region by applying:
\begin{equation}
P_{\rm up/low} = \sum_0^n 0.159^n\cdot (1-0.159)^{N-n}\cdot \frac{N!}{n!(N-n)!},
\end{equation}
where $N=100$ is the total number of data points. The summed terms are the probabilities for each data point and their combination formula while the summation from $0$ to $n$ includes all cases as extreme as or more extreme than the observed case (a summation from $0$ to $N$ results in 1). On the other hand, if there are $n\geq16$ points in the upper or lower region, we calculate the likelihood of having $\geq n$ points in that region
\begin{equation}
P_{\rm up/low} = \sum_n^N 0.159^n\cdot (1-0.159)^{N-n}\cdot \frac{N!}{n!(N-n)!}.
\end{equation}
Then, the joint likelihood or the p-value for a result equally or more extreme than the compiled dataset is given by
\begin{equation}\label{eq:P-value}
P=P_{\rm up}\cdot P_{\rm low}.
\end{equation}

\subsubsection{Model a23m150: the canonical IMF with $m_{\rm up}=150~M_\odot$}\label{sec:Model a23m150}

In Fig.~\ref{fig: RSMmaxMecl2} there are $0$ data points in the upper region and $34$ in the lower region. The resulting p-value is $P=2\cdot 10^{-13}$, that is, it is extremely unlikely that the real distribution is given by the synthetic clusters. The corresponding rejection confidence of the stochastic sampling hypothesis is $7\sigma$ (calculated in Section~\ref{sec:Confidence level of rejecting stochastic sampling}).

\subsubsection{Model a23m100: the canonical IMF slope with $m_{\rm up}=100~M_\odot$}\label{sec:Model a23m100}

\begin{figure}[!hbt]
    \centering
    \includegraphics[width=9cm]{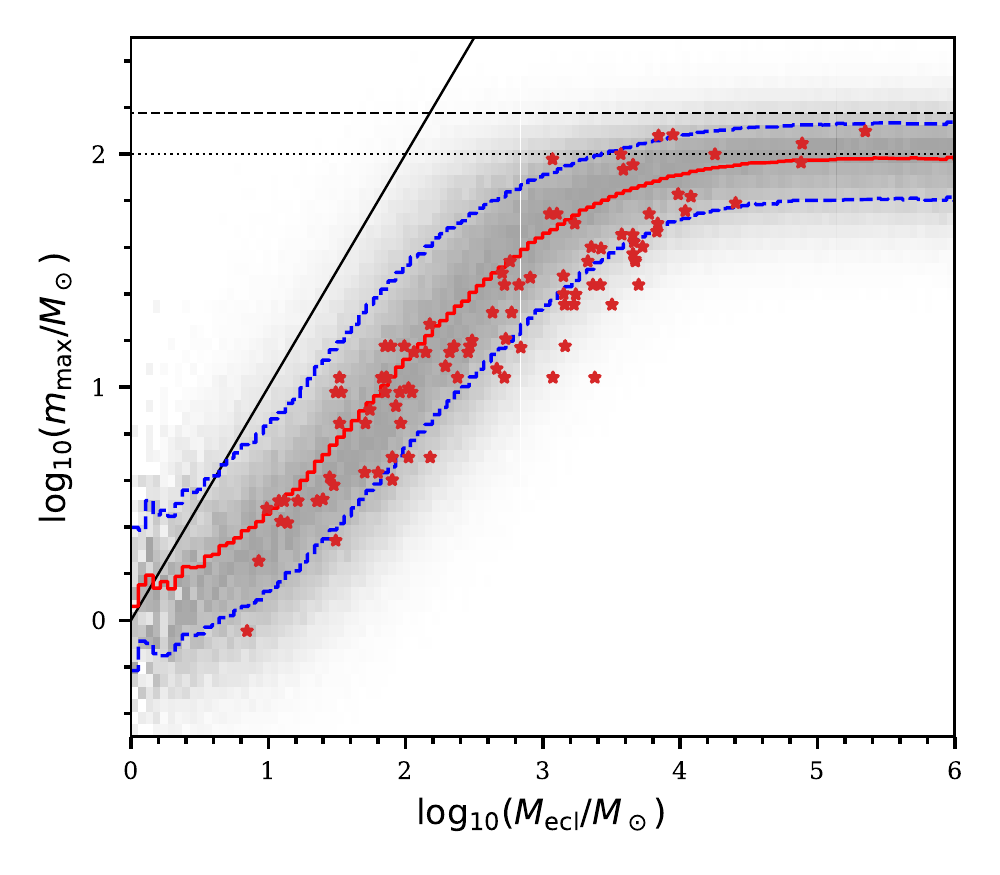}
    \caption{Same as Fig.~\ref{fig: RSMmaxMecl2} but for the a23m100 IMF model (Table~\ref{table: model names}). The additional horizontal dotted line denotes $m_{\rm up}=100~M_\odot$.}
    \label{fig:MmaxMeclRS2_MU_100}
\end{figure}

As mentioned in Section~\ref{sec: Method}, the uncertainty of the assumed IMF shape need to be taken into account. We first search for the stellar upper mass limit, $m_{\rm up}$, in Eq.~\ref{eq:xi} which improves the fit for $M_{\mathrm{ecl}}>10^4~M_\odot$ as is shown in Fig.~\ref{fig:MmaxMeclRS2_MU_100}. The best-fit value (with a mass changing step of $10~M_\odot$) is $m_{\rm up}=100~M_\odot$. There are $3$ and $25$ data points in the upper and lower regions, respectively, and the p-value increases to about $5\cdot 10^{-7}$. This corresponds to a $4.7\sigma$ confidence level to reject the stochastic sampling hypothesis (see Section~\ref{sec:Confidence level of rejecting stochastic sampling}).

\subsubsection{Model a268m130: non-canonical IMF}\label{sec:Model a268m130}

\begin{figure}[!hbt]
    \centering
    \includegraphics[width=9cm]{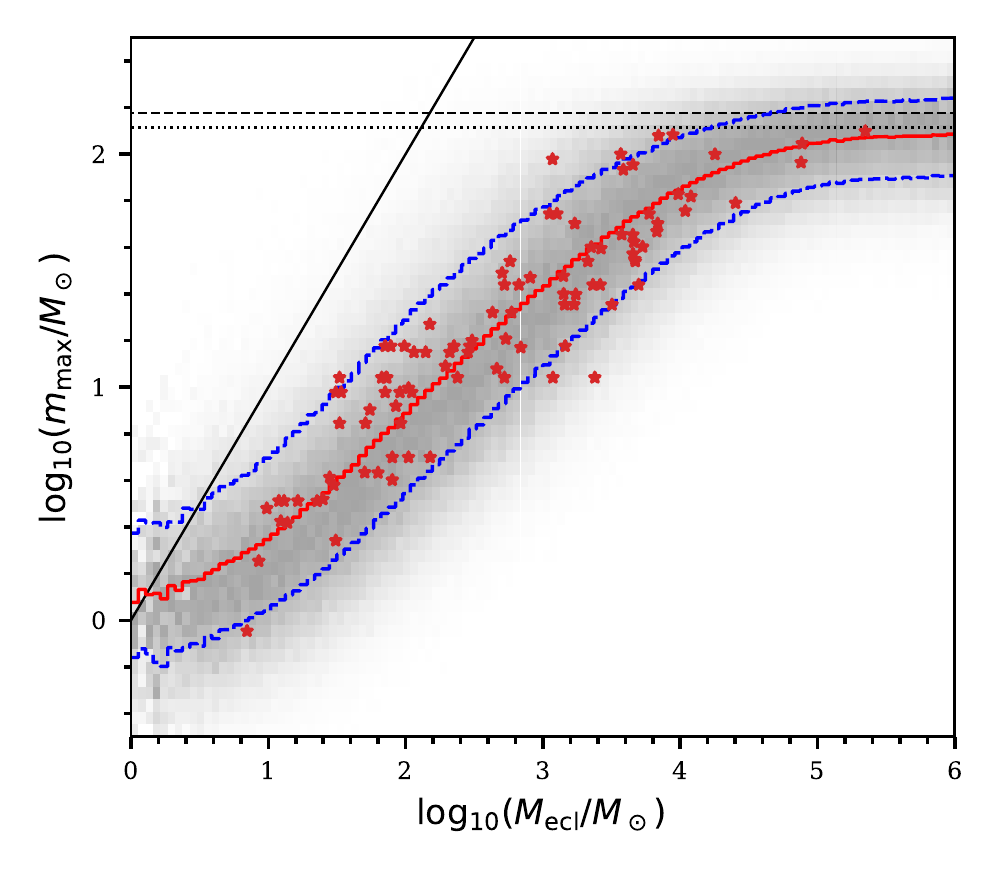}
    \caption{Same as Fig.~\ref{fig: RSMmaxMecl2} but for the a268m130 IMF model (Table~\ref{table: model names}) which best fits with the observed $m_{\rm max}$--$M_{\mathrm{ecl}}$ relation. The additional horizontal dotted line denotes $m_{\rm up}=130~M_\odot$.}
    \label{fig:MmaxMeclRS2_MU_130_a3_269}
\end{figure}

We also tested steeper high-mass IMF slopes. 
Here we allow $m_{\rm up}$ and $\alpha_3$ to vary using a grid with a $10~M_\odot$ step for different $m_{\rm up}$ and a 0.01 step for different $\alpha_3$. The best-fit combination for the observation is with $m_{\rm up}=130~M_\odot$ and $\alpha_3=2.68$ (model a268m130) in which case the data has six points in both upper and lower regions as shown in Fig.~\ref{fig:MmaxMeclRS2_MU_130_a3_269}. The p-value given by Eq.~\ref{eq:P-value} for this model is still less than $5.7 \cdot 10^{-6}$. Since the model fits the data nicely and already achieves the maximum equal number (six) of data points in the upper and lower regions, we can safely conclude that there is no room to further increase the p-value by adjusting IMF parameters for the applied dataset and test statistics. The obstacle to having a higher p-value is that the intrinsic variation of the stochastically sampled distribution is too large.

In addition, the probability of the adjusted $\alpha_3$ should be considered given $\alpha_3=2.3\pm 0.7$
for young star clusters \citep{2001MNRAS.322..231K}.
Assuming a normal distribution for the uncertainty of $\alpha_3$, this results in a probability for $\alpha_3\geq2.68$ to be about $29\%$.
Therefore, the joint p-value reduces to $5.7 \cdot 10^{-6}\times 29\%=1.7 \cdot 10^{-6}$. The corresponding rejection confidence of the stochastic sampling hypothesis is $4.5\sigma$ (see Section~\ref{sec:Confidence level of rejecting stochastic sampling}).

\subsection{Confidence level of rejecting stochastic sampling}\label{sec:Confidence level of rejecting stochastic sampling}

Here we estimate the confidence level of rejecting the stochastic sampling hypothesis. According to the Bayesian theory\footnote{$P(A|B) = \frac{P(B|A) \cdot P(A)}{P(B)}=\frac{P(B|A) \cdot P(A)}{P(B|A) \cdot P(A) + P(B|B) \cdot P(B)}$.}, two other estimations are needed: the prior probability of the null hypothesis, $Pr$, and the probability of observing a dataset less extreme than the real observation if the null hypothesis is not true, $R$. Together with the p-value estimated above, that is, the likelihood of observing a dataset as extreme as the real observation if the null hypothesis is true, $P$, we can have the probability of the null hypothesis being true given the observations, or in other words, the probability of wrongly rejecting the null hypothesis, a.k.a., the type I error, $A$:
\begin{equation}
A = \frac{Pr \cdot P}{Pr \cdot P + (1-Pr) \cdot (1-R)},
\end{equation}
with all parameters ranging from 0 to 1.

It is hard to estimate whether nature is influenced more by random processes or feedback self-regulation and the issue is still highly debated, thus, we assign both cases an equal prior probability (i.e. $Pr=0.5$). We also assume that half of the observation would be as extreme as the compiled dataset if some other sampling method instead of the stochastic sampling is true (i.e. $R=0.5$) since we do not have any information on whether the compiled dataset is more extreme than usual or not. With these values and $P\ll 1$, we have
\begin{equation}\label{eq: A}
A = \frac{P}{P + 0.5} \approx 2 P = 3.3 \cdot 10^{-6},
\end{equation}
with the p-values given by the a268m130 model in Section~\ref{sec: P-value test - stochastic sampling} that best reproduces the observed $m_{\rm max}$--$M_{\mathrm{ecl}}$ relation. The confidence level of rejecting the hypothesis is $1-A$. Given the cumulative normal distribution, this corresponds to a $4.5\sigma$ confidence level to reject the stochastic sampling hypothesis. If, on the other hand, the IMF is forced to have the canonical shape with $m_{\rm up}\geq150~M_\odot$, then the rejection is at more than $7\sigma$ confidence (Section~\ref{sec:Model a23m150}, cf. Section~\ref{sec:very massive stars} for the mass estimation of very massive stars).

\section{Optimal sampling}\label{sec: Test optimal sampling}

Optimal sampling is a deterministic sampling method that generates stellar masses which populate the IMF with no Poisson noise. The method is introduced in \cite{2013pss5.book..115K} to account for the small scatter of the observations. We refer the readers to \citet{2015A&A...582A..93S} and \citet{2017A&A...607A.126Y} for a further description of optimal sampling and the publicly available GalIMF code\footnote{https://github.com/Azeret/galIMF} that with stars optimally populates clusters and galaxies.

Optimal sampling assumes that the formation of stellar populations is perfectly self-regulated and, therefore, that a deterministic relationship exists between the mass of a star cluster and the mass of every single star within that cluster.
To sample the stellar masses, the IMF is divided into $N+1$ sections at mass limits of $x_{1}$ to $x_{N}$ (where $x_{i}>x_{i+1}$). These sections fulfil the mass conservation of the embedded star cluster
\begin{equation}\label{eq: M_tot split_str}
\begin{split}
M_{\rm ecl}&=
\int_{x_{\rm low}}^{x_{N}}m\xi(m)\,\mathrm{d}m + \int_{x_{N}}^{x_{N-1}}m\xi(m)\,\mathrm{d}m + ...\\ 
&+ \int_{x_{i+1}}^{x_i}m\xi(m)\,\mathrm{d}m + ... + \int_{x_2}^{x_1}m\xi(m)\,\mathrm{d}m\\
&=\int_{x_{\rm low}}^{x_{1}}m\xi(m)\,\mathrm{d}m,\\
\end{split}
\end{equation}
where $x_{\rm low}$ is the lowest possible stellar mass, $0.08~M_\odot$.
Each section gives exactly one star (therefore, the total number of stars is $N$):
\begin{equation}\label{eq: 1=_str}
1=\int_{x_{i+1}}^{x_i}\xi(m)\,\mathrm{d}m.
\end{equation}
In other words, if the integrated number of stars between two mass limits equals one, then the cluster must form exactly one star between these two mass limits with the optimally sampled stellar mass given by:
\begin{equation}\label{eq: M_i=_str}
m_i=\int_{x_{i+1}}^{x_i}m\xi(m)\,\mathrm{d}m.
\end{equation}
In addition, the value of $x_1$ fulfils
\begin{equation}\label{eq: optima_str}
1=\int_{x_1}^{x_{\rm up}}\xi(m)\,\mathrm{d}m,
\end{equation}
where the theoretical mass limit of the most massive star, $x_{\rm up}$, is set to $150~M_\odot$.
This set of equations can be solved analytically assuming a two-part power-law IMF \citep{2001MNRAS.322..231K} as is done in \citet{2017A&A...607A.126Y}.

\begin{figure}[!hbt]
    \centering
    \includegraphics[width=9cm]{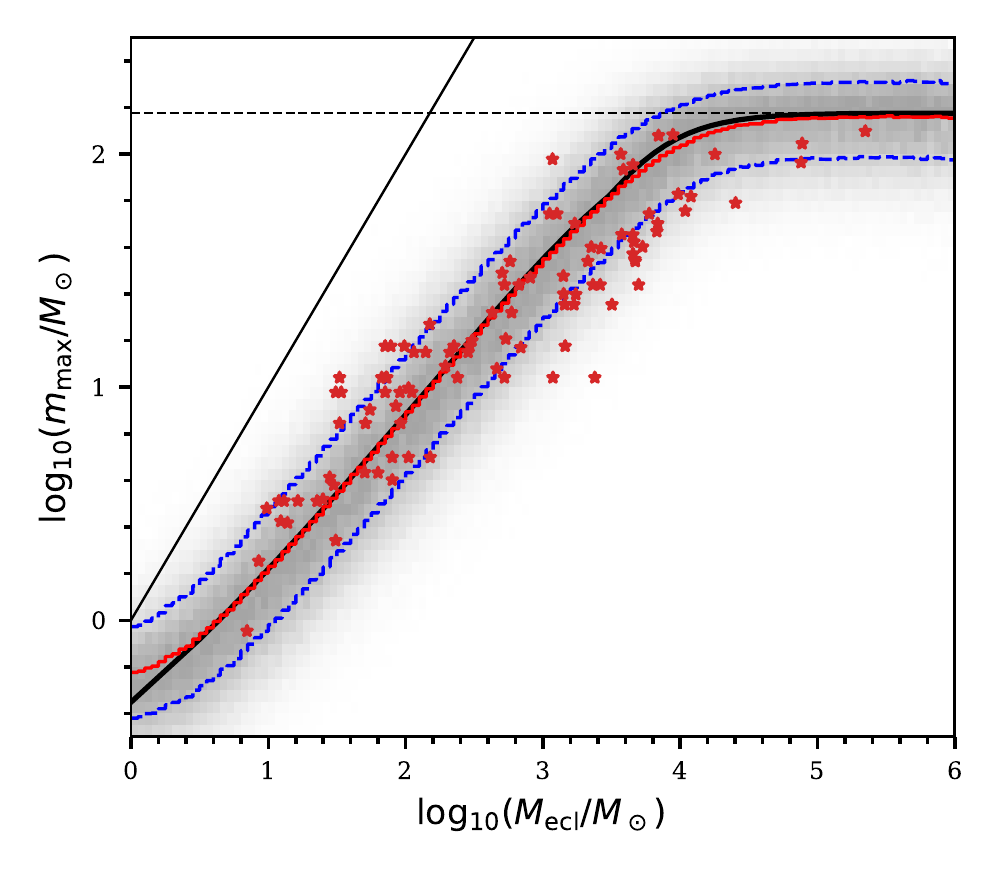}
    \caption{Same as Fig.~\ref{fig: RSMmaxMecl2} but with the optimally sampled clusters (the smooth black solid curve) with no intrinsic $m_{\rm max}$--$M_{\mathrm{ecl}}$ scatter. 
    By assuming a stellar population in an embedded cluster of stellar mass $M_{\rm ecl}$ in an optimal draw from the canonical IMF,  the observed distribution is obtained naturally once the observational uncertainties are taken into account. This is not the case for a randomly-drawn canonical IMF (Fig.~\ref{fig: RSMmaxMecl2}).
    }
    \label{fig:MmaxMeclRS2_opt}
\end{figure}

Applying the publicly available GalIMF code, here we sample stars from the canonical IMF\footnote{See ``example\_star\_cluster\_IMF.py'' in the GalIMF package. The code is also capable of optimally sampling stars from all the star clusters in a galaxy formed within a correlated 10~Myr star formation epoch \citep{2017A&A...607A.126Y} by implementing an empirical embedded cluster mass distribution, as is demonstrated by ``example\_galaxy\_wide\_IMF.py'' in the GalIMF package.}, that is, Eq.~\ref{eq:xi} with $\alpha_1=1.3$ and $\alpha_2=\alpha_3=2.3$ (see Section~\ref{sec: IMF bias} for IMF variation). The optimally sampled result, where the $m_{\rm max}$--$M_{\mathrm{ecl}}$ relation is deterministic, is shown in Fig.~\ref{fig:MmaxMeclRS2_opt}. 

We perform the same test as we applied for stochastic sampling in Section~\ref{sec: Test stochastic sampling}. There are 12 and 24 data points in the upper and lower regions, respectively. The resulting p-value is $0.004$, which is three orders of magnitude higher than the result of the stochastic sampling scenario. 
The total number of observations outside the standard deviation region ($12+24=36$) is only $13\%$ larger than the expected number of data in these regions for 100 data points ($100\times 31.8\%=32$). Adjusting the assumed IMF parameters (as is done for stochastic sampling in Section~\ref{sec: P-value test - stochastic sampling}) would be able to shift the synthetic $m_{\rm max}$--$M_{\mathrm{ecl}}$ relation and reach an equal expected number of points in the upper and lower regions and improve the fit.
From this, we conclude that optimal sampling is consistent with the observed $m_{\rm max}$--$M_{\mathrm{ecl}}$ relation and the star formation process appears to have a low level of randomness.

\section{Properties of the observed $m_{\rm max}$--$M_{\mathrm{ecl}}$ relation}\label{sec: Scatter of the data}

\subsection{The trend of the data}\label{sec:The trend of the data}

\begin{figure}[!hbt]
    \centering
    \includegraphics[width=9cm]{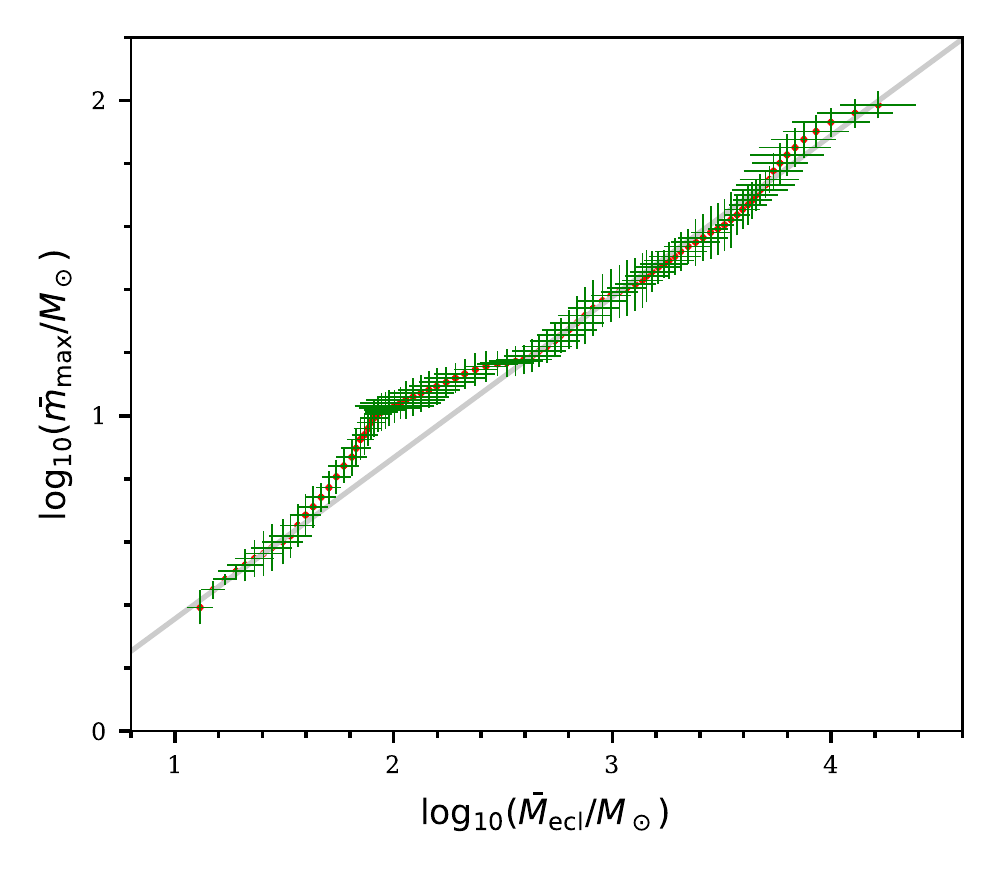}
    \caption{The average position of the observational data for the groups of 10 nearest points in the $m_{\rm max}$--$M_{\mathrm{ecl}}$ relation. The linear grey line highlights that the clusters with a mass between $10^{1.8}~M_\odot=63~M_\odot$ to $10^{2.6}~M_\odot=400~M_\odot$ depart from the linear relation.}
    \label{fig:scatter_2D}
\end{figure}

We visualise the trend of the observational data distribution (Fig.~\ref{fig: RSMmaxMecl}) by measuring the mean values of nearby points in both vertical and horizontal directions. The clusters are ranked and grouped by the nearest $m_{\rm max}$ (or $M_{\mathrm{ecl}}$) values, each group has 10 data points and the neighbouring group differs by only 1 data point. The average values for each group ($\bar{m}_{\rm max}$ or $\bar{M}_{\mathrm{ecl}}$) are calculated, paired monotonically, and shown in Fig.~\ref{fig:scatter_2D}. The error bar of the point is the standard deviation of the mean, $\sigma_{\rm mean}=\sigma_{10}/\sqrt[]{10}$, where $\sigma_{10}$ is the standard deviation of $m_{\rm max}$ (or $M_{\mathrm{ecl}}$) for a group of 10 data points.

Fig.~\ref{fig:scatter_2D} shows a feature above the grey line, which may relate to the stellar and star cluster formation physics. The flatter part of the trend around $m_{\rm max} = 13~M_\odot$ indicates that the formation of more massive stars may be inhibited at this mass, requiring a more massive star cluster to make the formation possible.

\subsection{The scatter of the data}

\begin{figure}[!hbt]
    \centering
    \includegraphics[width=9cm]{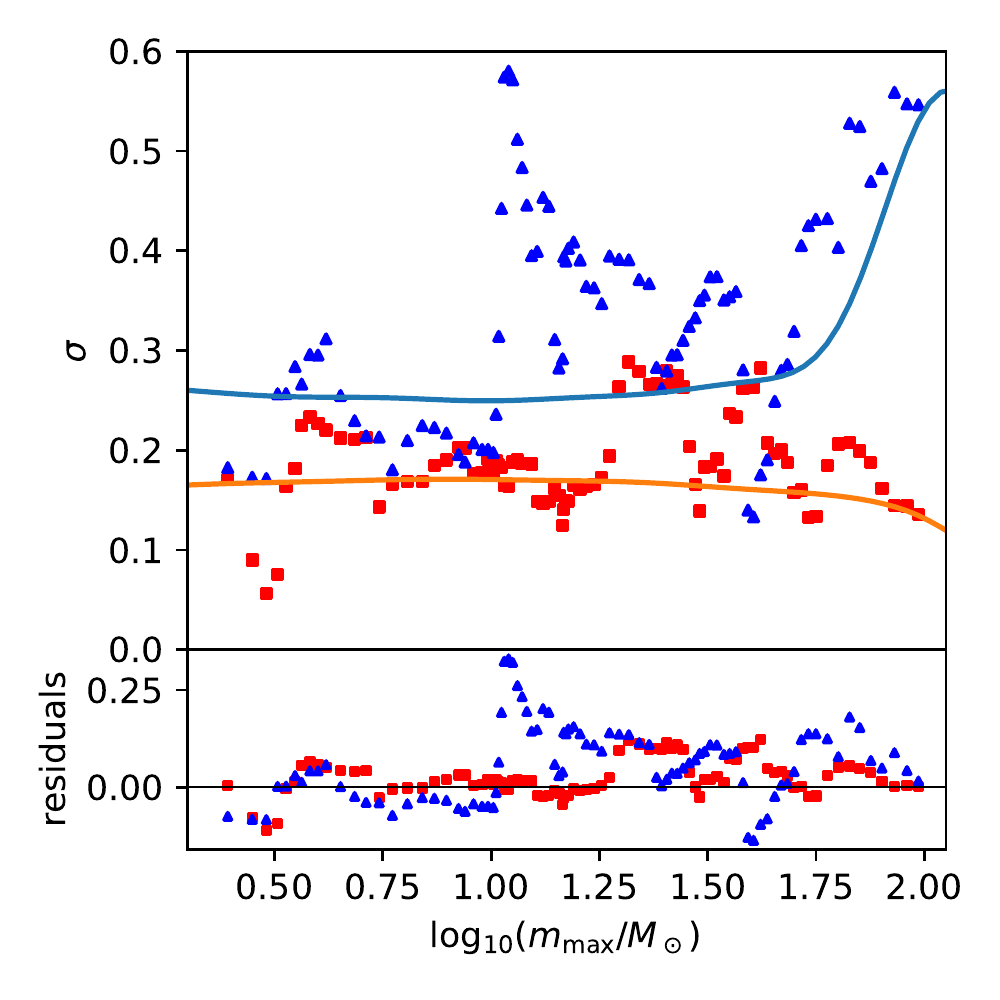}
    \caption{The standard deviation of $M_{\mathrm{ecl}}$ (blue triangles) and $m_{\rm max}$ (red squares) for the groups of 10 nearest points in the $m_{\rm max}$--$M_{\mathrm{ecl}}$ relation. The orange and the blue curves in the upper panel are the synthetic relations assuming optimal sampling of stars. The lower panel shows the difference between the model and the observations.}
    \label{fig:scatter_1D}
\end{figure}

The standard deviation, $\sigma_{10}$, of the nearest 10 data points for $m_{\rm max}$ and $M_{\mathrm{ecl}}$ (see Section~\ref{sec:The trend of the data}) is plotted for a given $m_{\rm max}$ in Fig.~\ref{fig:scatter_1D}.
In addition, the average variances of $10^4$ realisations of mock observations of 100 star clusters generated with the optimal sampling hypothesis (as described in Section~\ref{sec: Test optimal sampling}) are shown as solid curves.

The $M_{\mathrm{ecl}}$ values of the grouped data points show an increased variances at $m_{\rm max}\approx 13~M_\odot$ and $100~M_\odot$. The increase at about $100~M_\odot$ indicates a possible physical upper limit of stellar masses, $m_{\rm up}$. The residual of the observation and the optimal sampling model is shown in the lower panel of Fig.~\ref{fig:scatter_1D}. 
The signal to noise ratio (SNR\footnote{Here defined as 
${\rm SNR} = \left(\frac{\rm MAX[ \rm ABS[residual] ]}{\rm AVE\left[\rm ABS[residual]\right]}\right)^2$,
where $\mathrm{MAX}[x]$, $\mathrm{AVE}[x]$, and $\mathrm{ABS}[x]$ calculate the maximum, the average, and the absolute value of $x$, respectively.}) of the $M_{\mathrm{ecl}}$ residual peak at $m_{\rm max}\approx 13~M_\odot$ is 13.

The reason for the large $m_{\rm max}$ scatter above $10~M_\odot$ may be due to the photodissociation radiation produced by these massive stars such that accretion and growth to larger masses become difficult. \citep{2002ApJ...569..846Y,2007ARA&A..45..481Z}.
For example, \cite{1998MNRAS.298...93B} suggest a collisional build-up mechanism for forming massive stars ($>10~M_\odot$) which requires dense initial conditions that only exists in star clusters massive enough. Thus, they predict that the different mechanisms of forming stars may display
some structure at $\approx 10~M_\odot$. 
It is also possible that stellar-dynamical encounters eject the most massive stars for embedded clusters with masses in the range between $63~M_\odot$ to $400~M_\odot$ (cf. \citealt{2019MNRAS.484.1843W}). 
Indeed, \cite{2018A&A...612A..74K} find that the star clusters which are still building up and reach this mass range start forming massive stars but also eject them out of the cluster on a $<{\rm Myr}$ timescale, resulting in a zigzag evolution track towards the upper right in the $m_{\rm max}$--$M_{\mathrm{ecl}}$ relation, possibly similar to Fig.~\ref{fig:scatter_2D}.
Both scenarios suggest significant self-regulation in the star formation process.

\section{Uncertainty analyses}\label{sec: Uncertainty analyses}

Although the test in Section~\ref{sec: Test stochastic sampling} rejects the stochastic sampling hypothesis with a $4.5\sigma$ confidence level (allowing the IMF to deviate from its canonical shape) and at more than $7\sigma$ for the canonical IMF shape, the conclusion depends on the observational uncertainty estimations and data selection bias. This section discusses the effects that may compromise the rejection. 

It is worth noting that any uncertainty or hidden variable that is not accounted for (thus resulting in an underestimated mass measurement uncertainty) increases the scatter of the data points mimicking stochasticity which in truth is not present. Only an overestimated uncertainty may compromise the rejection of the stochastic sampling hypothesis.

\subsection{Selection bias}\label{sec: selection bias}

It is possible that star clusters with extreme properties, if they do exist, are pre-excluded from the sample, for example, single isolated stars or clusters consisting of only low-mass stars. This potential observational bias would make the resulting $m_{\rm max}$--$M_{\mathrm{ecl}}$ relation tighter than it really is, which mimics a more regulated star formation process.
Even if such clusters are not intentionally excluded, the freak star clusters can be more challenging to observe. A star cluster consisting of only low-mass stars would have a lower luminosity than a normal cluster with the same mass. On the other hand, if a star cluster has an $m_{\rm max}$ much higher than the value given by the average $m_{\rm max}$--$M_{\mathrm{ecl}}$ relation, by chance, it is unlikely that the second most massive star in this cluster has a similar mass as its most massive star. This means if the most massive star is missed or wrongly measured in the observation due to, for example, dynamical ejection, stellar evolution, or extinction, the extreme $m_{\rm max}$ value will also be missed. While for normal star clusters, even if the most massive star is wrongly estimated or missed, it is expected by both sampling methods that there are a few stars with lower but similar masses, reducing the measurement error of $m_{\rm max}$.

However, it is likely that freak star clusters do not exist. \citet{2017ApJ...834...94S} select, in particular, the seven most likely isolated massive stars in the entire LMC, each being distant from any known astronomical sources, and followed them up with deeper HST observations. In all these seven cases, a large young stellar population is detected in their immediate environments, which are distributed in a clustered fashion such that the $m_{\rm max}$--$M_{\mathrm{ecl}}$ relation is obeyed. 
In addition, dynamical ejection \citep{2011A&A...535A..29G,2011A&A...525A..17G,2012MNRAS.424.3037G,2012ApJ...746...15B,2012MNRAS.426.1416B,2016A&A...590A.107O}, including the two-step ejection process \citep{2010MNRAS.404.1564P}, is known to create apparently isolated massive stars.
Therefore, it is likely that all allegedly isolated massive stars are associated with normal star clusters with not good enough observations and should be excluded from our analyses. 
The stellar associations with a small number of stars have so far been confirmed, all having a total mass that is in agreement with the non-trivial $m_{\rm max}$--$M_{\mathrm{ecl}}$ relation \citep{2012ApJ...745..131K}.

\subsection{Bias from IMF assumption}\label{sec: IMF bias}

For many clusters in the sample, $M_{\mathrm{ecl}}$ is estimated by extrapolating the mass of the observed stars assuming a certain IMF. Considering the extreme case if only the mass of the most massive star is measured for star clusters. Then the cluster mass would be calculated based only on $m_{\rm max}$ which results in a deterministic $m_{\rm max}$--$M_{\mathrm{ecl}}$ relation without any scatter. We try to minimize this effect by excluding clusters with a small number of star detection but the effect cannot be fully avoided.
For example, the mass of NGC 6611 is estimated based on stars between 6 and $12~M_\odot$, assuming they account for a certain fraction of the cluster mass derived with a given IMF \citep[their section 5]{2007AJ....133.1092W}. 
The bias introduced by the IMF assumption depends on the completeness and the mass range of the observed stars, the applied IMF uncertainty, and whether additional $M_{\mathrm{ecl}}$ estimations with independent methods (dynamical estimation or from total luminosity) are taken into account. A more homogeneous study would help to better constrain and model this bias.

On the other hand, the shape of the sub-solar IMF depends on the metallicity of the stars (\citealt{2002Sci...295...82K,2013ApJ...771...29G,2018ApJ...855...20G,2019A&A...626A.124M,2020ARA&A..58..577S,2020A&A...637A..68Y,2022MNRAS.514.3660P}, and \citealt{Li2022}). Since the observed star clusters do not have the same metallicity, this introduces additional scatter to the expected $m_{\rm max}$--$M_{\mathrm{ecl}}$ relation, favouring the optimal sampling hypothesis.

\subsection{Contamination}

Field star contamination is inevitable. Studies employing dynamical and chemical tagging will further reduce the contamination in upcoming surveys.

\subsection{Mass uncertainties}\label{sec: Mass uncertainties}

\subsubsection{Uncertainties from Kirk \& Myers}\label{sec: Kirk Myers}

Star clusters with more than 20 detected members \cite[their table~1]{2011ApJ...727...64K} are adopted.
The uncertainty of $M_{\mathrm{ecl}}$ from \cite{2011ApJ...727...64K} is set to be the same as the uncertainty of $m_{\rm max}$ (which is roughly estimated to be $\pm 50\%$ by Kirk and Myers). 
\citet{2012ApJ...745..131K} find that massive stars can be better described if they assume an older cluster age (2 Myr instead of 1 Myr). This suggests that stars with different masses in the same star cluster may have different average ages and a single age assumption may introduce bias to the mass estimation of these pre-main-sequence stars. See also STARFORGE simulations where more massive stars in a star cluster form at a later time \citep{2022arXiv220510413G}. 
In addition, the statistical uncertainty of individual stars due to different extinction estimation errors and multiplicity are neglected. 
An extinction map based on stellar reddening is provided in \cite{2011ApJ...727...64K} but the uncertainty is unknown. 
A potential multiplicity of the stars could lead to a $20\%$ lower $m_{\rm max}$ and a $60\%$ higher $M_{\mathrm{ecl}}$ than estimated if every star in a cluster is an unresolved equal mass binary. 
Furthermore, it is possible that identified star cluster members are either contaminated by field stars or incomplete because of dynamical evolution and that some stars have not formed by the time of observation. 

\subsubsection{Uncertainties from Stephens et al.}

Clusters in \citet[their table~6]{2017ApJ...834...94S} are included. We adopt the mean cluster mass calculated with cluster ages of 1 and 2.5 Myr.
The uncertainties of $M_{\mathrm{ecl}}$ consider only the rough age uncertainty of the star cluster when fitting the cluster isochrone. The statistical and/or systematic uncertainties of extinction, distance, the grid of stellar models, etc. are not taken into account. Therefore, this uncertainty can be considered as a lower limit.
The adopted $m_{\rm max}$ uncertainty of about $40\%$ is estimated due to a large number of uncertain effects from spectral type model, multiplicity, extinction, and stellar age, none of which have been well estimated.

\subsubsection{Uncertainties from WKP13}\label{sec: WKP13}

A large number of star clusters is readily compiled in WKP13 (their table~A1). We select clusters with younger ages and a large number of detected stars, as described in Section~\ref{sec: data}.
According to WKP13, the uncertainty of $m_{\rm max}$ is estimated assuming a half spectral sub-class uncertainty of the most massive star, neglecting potential multiplicity. A physical upper stellar mass limit of $150~M_\odot$ is implemented in WKP13. More massive stars might be unresolved multiple stars or stars formed through dynamical mergers \citep{2012MNRAS.426.1416B,2018MNRAS.481..153O}. This upper mass limit would not affect our conclusion as discussed in the following section.

The uncertainty of $M_{\mathrm{ecl}}$ is mostly constituted by the assumption made in WKP13 that all stars could be unresolved binaries and that $50\%$ of the stars may be misidentified as cluster members, concluding a mass uncertainty of $\approx\pm 0.3~{\rm dex}$. This leads to a much higher average uncertainty of $M_{\mathrm{ecl}}$ than $m_{\rm max}$.
In reality, binaries would only lead to a $64\%$ to $68\%$ increase of the estimated system mass for stars between about $0.4$ and $55~M_\odot$ due to the non-linear relation between the stellar mass and luminosity with no increase for extremely massive stars due to the linear relation between the stellar mass and luminosity (see Eq.~\ref{eq:stellar mass-luminosity relation} below). That is, a less than $68\%$ increase of $M_{\mathrm{ecl}}$ instead of $100\%$ if all observed stars (mostly massive stars are observed) are unresolved equal mass binaries. The effect of field star contamination on $M_{\mathrm{ecl}}$ may also be lower than previously estimated. Massive stars contribute a large fraction of the cluster mass and they are less likely to be missed or contaminated than low-mass stars.

On the other hand, other sources of uncertainties, for example, IMF variation, variable extinction, stellar variability, gas expulsion and dynamical evolution, and stellar models, are not fully taken into account and the uncertainties in WKP13 are given as a lower limit. Therefore, it is possible that the adopted uncertainty is realistic.

\subsubsection{The masses of very massive stars}\label{sec:very massive stars}

The possible mass of the most massive star is still under debate, ranging from 100 to about $300~M_\odot$. \citet{2004MNRAS.348..187W}, \citet{2005Natur.434..192F}, \citet{2005ApJ...620L..43O}, and \citet{2006MNRAS.365..590K} suggest an upper limit to the initial masses of stars of about $150~M_\odot$ (the canonical value), while \cite{2010MNRAS.408..731C,2016MNRAS.458..624C} argues that the limit may be higher given the large systematic uncertainties in the stellar models of the massive stars. Other than the error from extinction and distance measurements, massive stars have a strong mass loss and a thick atmosphere difficult to model. Stellar simulations give very different evolution tracks by adopting a different amount of overshooting, rotation, mass loss rate, metallicity, and binary evolution. 
The uncertainties are larger for more massive stars \citep{2015ASSL..412....9M}. For example, the evolution tracks of \cite{2012A&A...537A.146E} failed to reproduce the luminosity and temperature of the stars in the Arches cluster. \cite{2013MNRAS.433.1114Y} note that the very massive stars may have luminosities during the advanced phases of their evolution similar to stars with initial masses below $120~M_\odot$, making these stars difficult to distinguish. This leads to the possibility of underestimating the $m_{\rm max}$ of the Arches cluster. 
\citet{2010MNRAS.408..731C,2016MNRAS.458..624C} revisit younger star clusters considering the line-blanketed model and give a higher mass estimation of the stars than WKP13, doubling the $m_{\rm max}$ value in the case of R136 in 30~Doradus. Applying different stellar evolution models, the mass of R136a1 (i.e. the most massive star in R136) estimated in \citet{2022arXiv220211080B} greatly exceeds the estimation in \citet{1998ApJ...493..180M}. 
On the other hand, the crowded cluster R136 suffers a limited photometric resolution and the very massive stars may turn out to be unresolved multi-stellar systems, as is the case in Pismis~24-1 \citep{2007ApJ...660.1480M}. Severe crowding within the core of R136 leads to larger measurement uncertainties and reduces the mass estimations \citep{2017IAUS..329..131R,2022ApJ...935..162K}. Indeed, stars less massive than about $20~M_\odot$ are not detected in the central 0.15 pc radius of R136 due to crowding. It is likely that more Wolf-Rayet stars, for example, with $50~M_\odot$ or $100~M_\odot$, exist than those that have been identified and these stars contribute to the luminosity of unresolved stellar systems. With a linear correlation between the mass and luminosity for the most massive stars, unresolved systems can significantly increase the stellar mass measurements. Even if the stars are indeed single stars, the strong stellar wind and highly uncertain mass-loss rate of very massive stars \citep{2022MNRAS.516.4052H} make it impossible to distinguish initial stellar masses above $200~M_\odot$ on the Hertzsprung–Russell diagram. The wind also reduces the mass of very massive stars to about $150~M_\odot$ within 1.6 Myr. Therefore, it is likely that a few stars with current masses exceeding the limit of $150~M_\odot$ but an age similar or older than 1.6 Myr (e.g. R136a1 as discussed in \citealt{2010MNRAS.408..731C}) form through mergers of massive binaries (\citealt{2012ApJ...746...15B,2012MNRAS.426.1416B,2018MNRAS.481..153O}, cf. \citealt{2018MNRAS.481..153O}). Given these considerations, the initial masses of very massive stars are still uncertain.

However, a different mass estimation of very massive stars would not significantly affect the present study because the scatter of the $m_{\rm max}$--$M_{\mathrm{ecl}}$ relation is mainly determined by star clusters with a mass below $8000~M_\odot$ (including 90\% of the clusters). 
The discussion if the birth masses of the most massive stars can be larger than the canonical $150~M_\odot$ value are relevant for massive star-burst clusters, such as R136, within which stellar mergers are common. But such clusters are very rare. Therefore, excluding these few data points would affect the best-fit $m_{\rm up}$ value (e.g. Section~\ref{sec:Model a268m130}) but not our conclusions.

\subsubsection{Correlations between mass estimation errors}

The two mass errors are not independent and there are three types of error correlations in general.
\begin{itemize}
	\item Systematic stellar mass error: The cluster mass is often calculated based on the observed stellar mass, which is extrapolated to the full stellar mass range with an assumed IMF. In this case, stellar model, cluster age, distance, and average dust extinction estimation errors would affect the estimation of both stellar and star-cluster masses with a positive correlation.
	\item Statistic stellar mass error: random measurement error for $m_{\rm max}$ contributes to the $M_{\mathrm{ecl}}$ error. For example, very massive stars have a large uncertainty of their evolution and the observational parameters due to a strong mass loss and probable binary evolution \citep{2005MNRAS.357.1088Z}.
	However, the error correlation is insignificant when the $m_{\rm max}/M_{\mathrm{ecl}}$ ratio is low, which is the case for the majority of the stochastically sampled star clusters. 
	There are only about $10\%$ of the star clusters (stochastically sampled from the canonical IMF) that have $m_{\rm max}/M_{\mathrm{ecl}}>0.3$ and about $4\%$ with $m_{\rm max}/M_{\mathrm{ecl}}>0.5$.
	\item Unresolved multi-stellar system: If a multi-stellar system is unresolved, the mass estimation of the primary star is higher than the correct value while the mass estimation of the star cluster is lower than the correct value, therefore, the errors are anti-correlated.
\end{itemize}
The values of different types of errors relate differently and non-linearly with the mass estimations of stars and clusters. Ideally, one simulates realistic statistics and systematic uncertainties originating from the stellar model, age estimation, distance, extinction, multiplicity, etc., with detailed forward modelling. However, there are practical difficulties in doing so.
Firstly, the uncertainty analyses in the literature are not comprehensive. That is, we do not have detailed information about different types of errors. Usually, only a roughly estimated total error is given if provided at all. Secondly, the measurements are adopted from different studies applying different methods of mass estimation with different uncertainties such that there is no uniform way to simulate mock uncertainties that can represent the dataset. Thirdly, the observational uncertainty of freak star clusters (Section~\ref{sec: selection bias}) cannot be estimated because they are not observed and only exist in the mock dataset.

\subsubsection{Non-Gaussian mass error distribution}

We test in addition the scenario that mock mass measurement errors follow a truncated normal distribution clipped to the range 0 to $2\sigma$. Again, the upper and lower variances are different (see Section~\ref{sec: data}), therefore, the upper and lower clip values are calculated separately. The change reduces the scatter of the synthetic $m_{\rm max}$--$M_{\mathrm{ecl}}$ relation and increases the p-value of our two-sample test. An equal number of data points in the upper and lower regions can increase to at most seven with slightly adjusted IMF parameters. The resulting confidence level for rejecting the stochastic sampling hypothesis is reduced to $4\sigma$ for the case allowing $m_{\rm up}$ and $\alpha_3$ to deviate from the canonical IMF, which is still significant.

\subsection{Conditions for stochastic sampling to reproduce the observed $m_{\rm max}$--$M_{\mathrm{ecl}}$ relation}

Considering that the $M_{\mathrm{ecl}}$ uncertainties are not well-estimated in much of the adopted literature, we explore what would be the correct mean $M_{\mathrm{ecl}}$ uncertainty if the stochastic sampling hypothesis is correct. We find that for the stochastically generated star clusters to reproduce the observed tight $m_{\rm max}$--$M_{\mathrm{ecl}}$ relation, there needs to be a high probability (>50\%) of omitting the most massive stars in our mock observation and that the applied cluster mass uncertainties must be much smaller than the average value of the observational dataset. Specifically, if the average uncertainty of $M_{\mathrm{ecl}}$ is the same as the average uncertainty of $m_{\rm max}$ on the logarithmic scale (i.e. about 0.15~dex instead of 0.3~dex), the resulting mock observations would have a similar scatter as the observed points.

However, Sections~\ref{sec: Kirk Myers} to \ref{sec: WKP13} demonstrate that the adopted mass uncertainties are often given as a lower limit. Therefore, these requirements are not likely to be fulfilled.

\section{Discussion}\label{sec: Discussion}

\subsection{Previous studies}

\label{point:scatter error} Reproducing the mean trend of the $m_{\rm max}$--$M_{\mathrm{ecl}}$ relation \citep{2013A&A...553A..31C,2021PASJ...73.1036H} is possible with both sampling hypothesis but the small scatter of the $m_{\rm max}$--$M_{\mathrm{ecl}}$ relation favours optimal sampling.
A possibly larger measurement uncertainty of the adopted dataset, as is suggested by \citet{2014PhR...539...49K} and sections~\ref{sec: Kirk Myers} to \ref{sec: WKP13} above, would conclude an even smaller intrinsic scatter of the $m_{\rm max}$--$M_{\mathrm{ecl}}$ relation and strengthen the argument favouring optimal sampling.

We note that a test between deterministic sampling and stochastic sampling (e.g. the present work) is different from a test between stochastic sampling and stochastic sampling with a cluster-mass-dependent truncation for the mass of the most massive star (e.g. \citealt{2013MNRAS.432.3097H}).
The earlier works studying the $m_{\rm max}$--$M_{\mathrm{ecl}}$ relation suggest the truncated stochastic sampling scenario (e.g. \citealt[their sorted random sampling]{2006MNRAS.365.1333W}) which is claimed to be ruled out by many studies (e.g. \citealt{2009A&A...495..479C}, \citealt{2011ApJ...741L..26F}, \citealt{2013ApJ...767...51A,2014ApJ...793....4A}, and \citealt{2017MNRAS.464.1738D}).
But these are not falsifications of optimal sampling, which is a pure deterministic sampling (Eqs.~\ref{eq: 1=_str} and \ref{eq: M_i=_str}, see also \citealt[their equation~4.9]{2013pss5.book..115K}, \citealt[their equations~6 and 7]{2015A&A...582A..93S}) and distinct from stochastic sampling under a deterministic truncation.
For example, \citet{2010MNRAS.401..275W} discuss the observational $m_{\rm max}$--$M_{\mathrm{ecl}}$ relation, therefore, their $m_{\rm max}$ corresponds to ``the mass'' of the most-massive star and the best-fit relation is consistent with the observation by definition. However, if the observational $m_{\rm max}$ is taken as ``the mass limit'' of the most-massive star for stochastic sampling, the stochastically sampled most massive star would be systematically less massive than $m_{\rm max}$, therefore, being inconsistent with the observations.
Similarly, according to the optimal sampling hypothesis, star clusters of $100$, $400$, and $1000~M_\odot$ and with ($1/10$th of) the solar metallicity must have their most-massive star being about $8~(9)$, $20~(22)$, and $36~(40)~M_\odot$, respectively (calculated by the open source GalIMF code, \citealt{2017A&A...607A.126Y}), which are in good agreement with the observation given by \citet{2012ApJ...749...20K}.

In summary, truncated statistical sampling is not directly related to the discussion of whether the star formation process is highly self-regulated. A significantly tight $m_{\rm max}$--$M_{\mathrm{ecl}}$ relation can and can only exist because of optimal sampling instead of truncated random sampling, which is in agreement with previous studies.

\subsection{Practical considerations}

The deterministic optimal sampling method is useful when one aims to study the evolution of an ``average'' stellar population closely resembling the IMF. It is also much more efficient in situations when only the masses of the most massive stars are of interest \citep{2020MNRAS.492....8A,2022MNRAS.509.5938H} because the numbers are deterministic and can be given analytically (Eqs.~\ref{eq: M_tot split_str} to \ref{eq: optima_str}).
While in the case of stochastic sampling, sampling only the massive stars, such as in \citet{2017MNRAS.466.1903G}, does not follow the probabilistic interpretation of the IMF where the formation of stars is an independent process. 
Therefore, one has to always first finish sampling all the stars in a star cluster before the most massive stars can be identified without bias (e.g. \citealt{2018MNRAS.481.2548G}). This can lead to a significant cost difference for high-resolution simulations.
Intrinsically, this is because the stochastic sampling hypothesis contradicts mass conservation. Given the total mass of a molecular cloud and a star formation efficiency (which may be universal), what is being constrained is the mass instead of the number of the stars that can form in this cloud. Macroscopic physics as we know it is not stochastic.
There are a variety of treatments regarding when to stop the random drawing of stars in order to approximate mass conservation to some level but mass conservation and reproducing the IMF cannot be simultaneously fulfilled with stochastic sampling \citep[their section 2]{2021MNRAS.502.5417S}. 

On the other hand, the optimal sampling hypothesis also faces a fundamental problem: the total mass of a real single formation event of a star cluster cannot be well-defined because stars did not form in one instance and there may be multiple populations of stars formed (or captured) in the same star cluster \citep{2007MNRAS.375..855P,2008IAUS..246...71P,2009MNRAS.397..488P,2018A&A...612A..74K,2020MNRAS.491..440W,2022MNRAS.516.3342W}.
In addition, optimal sampling cannot be directly applied to ``star particles'' that represent multiple star clusters or a fraction of a star cluster. This problem can be bypassed by stochastic sampling from a list of optimally sampled stellar masses of a galaxy (i.e. OSGIMF, \citealt{2017A&A...607A.126Y}) and may eventually be resolved by galaxy simulation models where the stellar particle represents the true cluster mass.

\subsection{Properties of the star forming clouds}\label{Appendix: Properties of the star forming clouds}

\begin{figure}[!hbt]
    \centering
    \includegraphics[width=9cm]{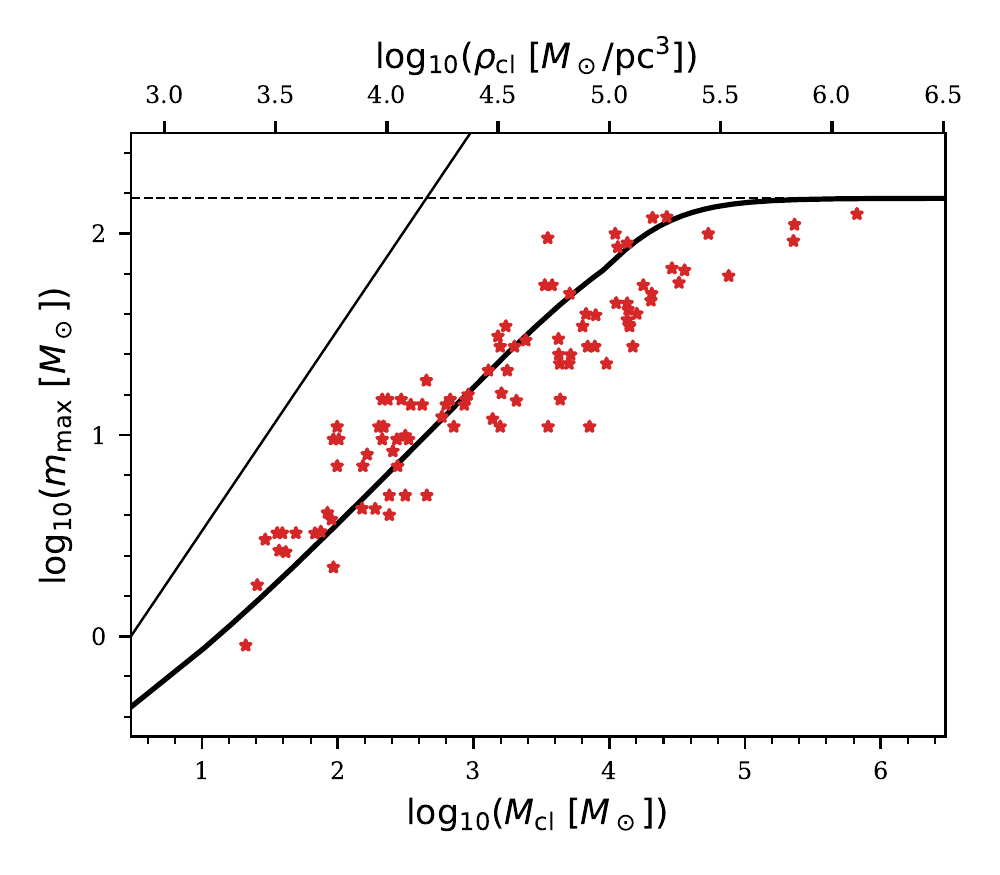}
    \caption{Same as Fig.~\ref{fig:MmaxMeclRS2_opt} but for the pre-cluster molecular cloud mass, $M_{\rm cl}$, and cloud core density, $\rho_{\rm cl}$, assuming a star formation efficiency of $33\%$ and Eq.~\ref{eq:rho_cl}. The thick curve is given by the optimal sampling hypothesis (Section~\ref{sec: Test optimal sampling}).}
    \label{fig:Mclmmax_opt}
\end{figure}

Assuming a star formation efficiency of $33\%$, the total mass of the pre-cluster molecular cloud, $M_{\rm cl}$, can be estimated. The scaling relation between $M_{\rm ecl}$ and the pre-cluster molecular cloud core density, $\rho_{\rm cl}$, from \citet[their eq. 7]{2017A&A...607A.126Y} is applied to estimate $\rho_{\rm cl}$, that is
\begin{equation}\label{eq:rho_cl}
\log_{10}(\rho_{\mathrm{cl}}~[M_\odot/{\rm pc}^3])=0.61\cdot \log_{10}(M_{\rm ecl}~[M_\odot])+2.85.
\end{equation}
Therefore, the $m_{\rm max}$--$M_{\mathrm{ecl}}$ relation results in the relations of these quantities as is shown in Fig.~\ref{fig:Mclmmax_opt}.

\begin{figure}[!hbt]
    \centering
    \includegraphics[width=9cm]{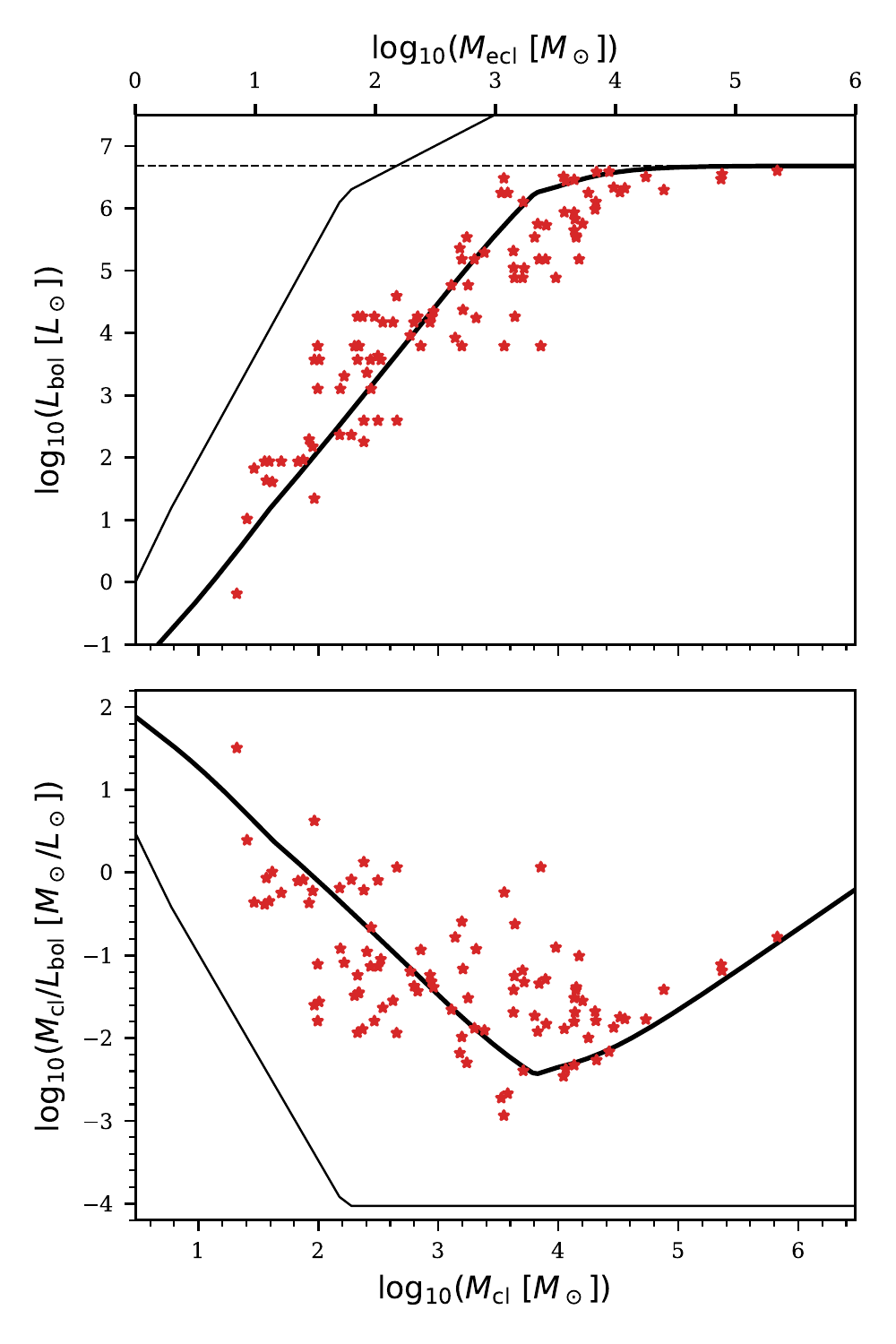}
    \caption{Same as Fig.~\ref{fig:MmaxMeclRS2_opt} and \ref{fig:Mclmmax_opt} but for the bolometric luminosity (upper panel) and the mass-to-light ratio (lower panel). The horizontal coordinates of the upper and lower panel are identical to Fig.~\ref{fig:MmaxMeclRS2_opt} and \ref{fig:Mclmmax_opt} ($M_{\rm cl}=3M_{\rm ecl}$).}
    \label{fig:MtoL_opt}
\end{figure}

In addition, the bolometric luminosity, $L_{\rm bol}$, of the most massive star is derived following the mass--luminosity relation summarised in \citet[their eq. 1]{2019A&A...629A..93Y}, following \citet[their chapter 1.3.8]{2004adas.book.....D} for the low-mass stars and \citet[their chapter 5.7]{2006essp.book.....S} for the stars more massive than about 2 M$_\odot$\footnote{\citet{2006essp.book.....S} argue that for more massive stars ($>20~M_\odot$), $L\propto M$ because radiation pressure is dominating the total pressure instead of gas pressure. However, the exact mass boundary to apply a linear $L$--$M$ relation is not provided. Observations for stars up to $50~M_\odot$ suggest a $L$--$M$ relation with a power-law index of about 2.8 instead of 1 (\citealt{2007AstL...33..251V} and up to $64~M_\odot$ according to \citealt[their Table 7]{2018MNRAS.479.5491E}). For simplicity, here we assume the widely applied power-law index of 3.5 for stars below $55.41~M_\odot$ and a linear $L$--$M$ relation for more massive stars, as is assumed in \citet{2019A&A...629A..93Y}.}
\begin{equation}\label{eq:stellar mass-luminosity relation}
    \frac{L_{\rm bol}}{L_\odot} =
    \begin{cases} 
    0.23\left(\frac{m}{M_\odot}\right)^{2.3}, & \frac{m}{M_\odot} < 0.421; \\
    \left(\frac{m}{M_\odot}\right)^{4}, & 0.421 \leq \frac{m}{M_\odot}<1.96; \\
    1.4\left(\frac{m}{M_\odot}\right)^{3.5}, & 1.96\leq \frac{m}{M_\odot}<55.41; \\
    32000\frac{m}{M_\odot}, & 55.41\leq \frac{m}{M_\odot},
    \end{cases}
\end{equation}
where $m$ is the stellar initial mass.
The derived $L_{\rm bol}$ values and the ratios between $M_{\rm cl}$ and $L_{\rm bol}$ are shown in Fig.~\ref{fig:MtoL_opt}.

\begin{figure}[!hbt]
    \centering
    \includegraphics[width=9cm]{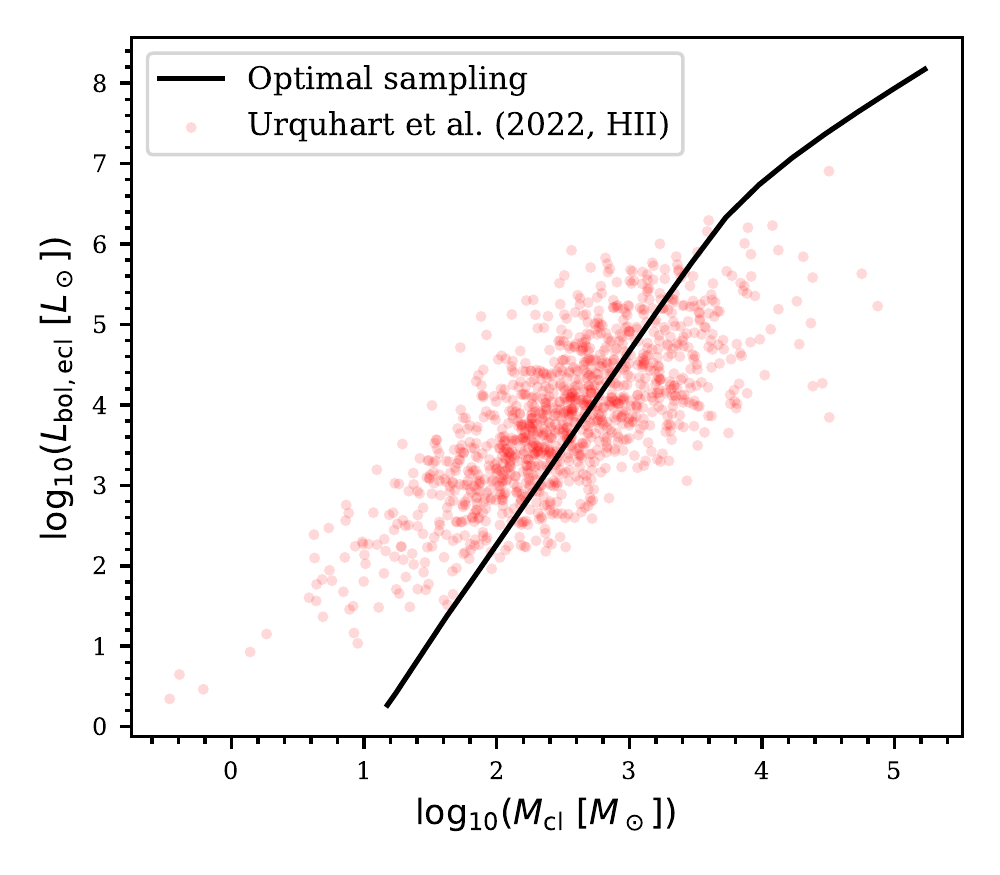}
    \caption{The total stellar bolometric luminosity--pre-cluster molecular cloud mass relation. The solid curve shows the prediction from the optimal sampling method, where the cluster mass relates to the mass of each star formed in that cluster deterministically, according to Eq.~\ref{eq: M_tot split_str} to \ref{eq: optima_str}. The total stellar luminosity is calculated with Eq.~\ref{eq:stellar mass-luminosity relation}. The red circles are the observational data adopted from \citet{2022MNRAS.510.3389U}, their H~II region related clumps.}
    \label{fig:total_Lbol_opt}
\end{figure}

Finally, the total combined bolometric luminosity of all the optimally-sampled stars in a young star cluster is calculated, assuming that the pre-main-sequence stars have a bolometric luminosity equal to the zero-age-main-sequence (ZAMS) stars given by Eq.~\ref{eq:stellar mass-luminosity relation}. The relation predicted by the optimal sampling hypothesis is compared with the observed mass, $M_{\rm fwhm}$, and bolometric luminosity within the FWHM source sizes of the star-forming dense molecular clumps \citep[their H II region related clumps]{2022MNRAS.510.3389U} in Fig.~\ref{fig:total_Lbol_opt}, assuming $M_{\rm cl} = M_{\rm fwhm}$.

The optimal sampling model is in qualitative agreement with the data (Fig.~\ref{fig:total_Lbol_opt}), which include complex radiation processes and a contribution from accretion luminosity. The model, in contrast, counts only the ZAMS luminosities of all stars in the embedded cluster assuming
them to appear instantly at the same time, while star-formation might proceed over a Myr (Section~\ref{sec: Kirk Myers}). 
In addition, massive stars can already be ejected out of the cloud within 1 Myr \citep{2018A&A...612A..74K}. These effects result in a lower observed luminosity than our model for massive clusters.
On the other hand, the observational data systematically become more luminous than the optimal sampling model towards smaller cloud mass $M_{\rm cl}$. This is likely a consequence of the following physics: at the massive end, most of the bolometric luminosity comes from the massive stars which are already on the main sequence when they become detectable in HII regions. The model thus well represents these data. Low-mass embedded clusters, on the other had, are predominantly populated by low-mass stars. These are, when they become visible, still contracting onto the main sequence with significantly larger luminosities when very young. It will be an interesting problem to include pre-main sequence evolutionary tracks into the optimal sampling model to test, in the future, if the observational data shown in Fig.~\ref{fig:total_Lbol_opt} better agree with such a model. At the low-mass cluster end, the optimal sampling model does not expect the formation of massive stars in these clusters while some observed low-mass clumps appear to have HII region features. The ionising radiation responsible for creating the HII region may be due to low-mass pre-main sequence stars producing a higher ionising flux than their ZAMS counterparts due to their flaring activity (e.g. \citealt{2020AN....341..519R}).

\section{Conclusions}\label{sec: Conclusion}

In this study, we demonstrate that the stochastic sampling hypothesis fails to reproduce the small scatter of the observed $m_{\rm max}$--$M_{\mathrm{ecl}}$ relation such that the hypothesis is rejected at a $4.5\sigma$ significance level if the measurement uncertainties of $m_{\rm max}$ and $M_{\mathrm{ecl}}$ are not overestimated, the compiled dataset is unbiased, and if the IMF is allowed to be adjusted. If, on the other hand, the IMF has the canonical form and $m_{\rm up}=150~M_\odot$, then the data is inconsistent with random sampling of stellar masses from the IMF as being a physical model with $7\sigma$ confidence.
If the adopted mass uncertainties are underestimated, which is the case when certain uncertainties are omitted, the conclusion becomes stronger.
In comparison, optimal sampling from an invariant canonical IMF and without an intrinsic dispersion of $m_{\rm max}$ values for a given $M_{\mathrm{ecl}}$ describes the observed $m_{\rm max}$--$M_{\mathrm{ecl}}$ relation naturally.

The optimal sampling model predicts a strong correlation between the bolometric luminosity of a young stellar population and the mass of the embedded cluster. We calculate this relation assuming a freshly hatched optimally sampled stellar population to be on the zero-age main sequence. The correlation agrees qualitatively well (Fig.~\ref{fig:total_Lbol_opt}) with the data from the ATLASGAL survey \citep{2022MNRAS.510.3389U}. Future modelling will likely improve the agreement once pre-main sequence stellar evolution tracks are incorporated into the evolution of the bolometric luminosity of low-mass stars.

In addition, we discover a large number of clusters with $m_{\rm max}\approx 13~M_\odot$. This may indicate that the formation of more massive stars is inhibited beyond $13~M_\odot$ and/or there is a strong dynamical ejection for the most massive stars for embedded clusters with stellar masses in the range of about 63 to $400~M_\odot$, supporting significant self-regulation in the star formation process.

A homogeneous and consistent observational survey of embedded clusters and their stellar content would help improve the reliability of our statistical test by providing a more reliable uncertainty analysis. Studies examining the properties of dwarf galaxies (as mentioned in Section~\ref{sec:intro}) will continue to provide us with constraints on the self-regulation in the star formation process. For example, if stars were stochastically sampled, then the integrated galaxy-wide IMF of many dwarf galaxies would be the same as the canonical IMF (cf. \citealt{2009ApJ...706..599L}), meaning that they would have the same supernova formation efficiency as massive galaxies (cf. \citealt{2021RAA....21..306Z}) and their chemical evolution would be understandable with an invariant canonical IMF (cf. \citealt{2020A&A...637A..68Y,2021NatAs...5.1247M}). See also \citet[their section~1.6.3]{2021arXiv211210788K} for a further discussion of these tests.

In conclusion, we still do not have a comprehensive theory on how stars form. Nature is deterministic, so star formation cannot be stochastic. But, since we do not know all variables (e.g. rotation, temperature) of the embedded-cluster forming molecular cloud clump, this lack of knowledge would transpire as an apparent stochasticity on top of the optimal sampling process such that the practical sampling method could be in between the statistical and optimal sampling extremes. The Python GalIMF code that performs optimal sampling for star clusters and entire galaxies is publicly available (see Section~\ref{sec: Test optimal sampling}).

\begin{acknowledgements}
Z.Y. acknowledges support from National Natural Science Foundation of China under grant number 12203021, the Jiangsu Funding Program for Excellent Postdoctoral Talent under grant number 2022ZB54, the Fundamental Research Funds for the Central Universities under grant number 0201/14380049, the National Natural Science Foundation of China under grants No. 12041305 and 12173016, the Program for Innovative Talents, Entrepreneurs in Jiangsu, the science research grants from the China Manned Space Project with NO.CMS-CSST-2021-A08 (IMF).
\end{acknowledgements}

\medskip
 
\bibliography{library}

\longtab[1]{
\begin{longtable}{ccccccc}
\caption{Literature data for the $m_{\rm max}$--$M_{\mathrm{ecl}}$ relation.}
\label{table: Literature data}\\
\hline\hline
No. & Designation & $\log_{10}(M_{\rm ecl}/M_\odot)$ & $\log_{10}(m_{\rm max}/M_\odot)$ & Age (Myr) & References \\
\hline
\endfirsthead
\caption{continued.}\\
\hline\hline
No. & Designation & $\log_{10}(M_{\rm ecl}/M_\odot)$ & $\log_{10}(m_{\rm max}/M_\odot)$ & Age (Myr) & References \\
\hline
\endhead
\hline
\endfoot
\hline
\endlastfoot
1	&	Taurus No.1 	& $	0.930	^{+0.176}_{-0.301}$ & $	0.254	^{+0.176}_{-0.301} $	&	1-2	&	1	\\
2	&	Taurus No.2 	& $	1.077	^{+0.176}_{-0.301}$ & $	0.512	^{+0.176}_{-0.301} $	&	1-2	&	1	\\
3	&	Taurus No.4 	& $	1.215	^{+0.176}_{-0.301}$ & $	0.512	^{+0.176}_{-0.301} $	&	1-2	&	1	\\
4	&	Taurus No.6 	& $	1.091	^{+0.176}_{-0.301}$ & $	0.425	^{+0.176}_{-0.301} $	&	1-2	&	1	\\
5	&	Taurus No.7 	& $	1.140	^{+0.176}_{-0.301}$ & $	0.418	^{+0.176}_{-0.301} $	&	1-2	&	1	\\
6	&	Lupus3 No.1 	& $	0.989	^{+0.176}_{-0.301}$ & $	0.480	^{+0.176}_{-0.301} $	&	1-2	&	1	\\
7	&	ChaI No.2 	    & $	1.355	^{+0.176}_{-0.301}$ & $	0.511	^{+0.176}_{-0.301} $	&	1-2	&	1	\\
8	&	ChaI No.3 	    & $	1.115	^{+0.176}_{-0.301}$ & $	0.512	^{+0.176}_{-0.301} $	&	1-2	&	1	\\
9	&	IC348 No.1 	    & $	1.702	^{+0.176}_{-0.301}$ & $	0.634	^{+0.176}_{-0.301} $	&	1-2	&	1	\\
10	&	No.3	& $	0.845	^{+0.331}_{-0.368}$ & $	-0.046	^{+0.125}_{-0.051} $	&	$2.6\pm0.8$	&	WKP13	\\
11	&	No.18	& $	1.398	^{+0.318}_{-0.319}$ & $	0.519	^{+0.115}_{-0.157} $	&	0.6	&	WKP13	\\
12	&	No.19	& $	1.447	^{+0.309}_{-0.333}$ & $	0.613	^{+0.095}_{-0.121} $	&	0.1	&	WKP13	\\
13	&	No.21	& $	1.477	^{+0.176}_{-0.301}$ & $	0.580	^{+0.184}_{-0.325} $	&	1	&	WKP13	\\
14	&	No.22	& $	1.491	^{+0.301}_{-0.315}$ & $	0.342	^{+0.038}_{-0.041} $	&	2	&	WKP13	\\
15	&	No.23	& $	1.491	^{+0.315}_{-0.315}$ & $	0.978	^{+0.101}_{-0.133} $	&	1	&	WKP13	\\
16	&	No.24	& $	1.519	^{+0.320}_{-0.314}$ & $	0.845	^{+0.133}_{-0.192} $	&	1	&	WKP13	\\
17	&	No.25	& $	1.519	^{+0.320}_{-0.314}$ & $	1.041	^{+0.135}_{-0.196} $	&	1	&	WKP13	\\
18	&	No.26	& $	1.531	^{+0.307}_{-0.327}$ & $	0.978	^{+0.101}_{-0.133} $	&	1	&	WKP13	\\
19	&	No.30	& $	1.708	^{+0.314}_{-0.327}$ & $	0.845	^{+0.133}_{-0.192} $	&	1	&	WKP13	\\
20	&	No.31	& $	1.740	^{+0.313}_{-0.309}$ & $	0.903	^{+0.097}_{-0.125} $	&	3	&	WKP13	\\
21	&	No.34	& $	1.799	^{+0.120}_{-0.166}$ & $	0.633	^{+0.166}_{-0.272} $	&	0.1	&	WKP13	\\
22	&	No.35	& $	1.826	^{+0.311}_{-0.321}$ & $	1.041	^{+0.135}_{-0.196} $	&	1	&	WKP13	\\
23	&	No.36	& $	1.851	^{+0.310}_{-0.320}$ & $	0.978	^{+0.101}_{-0.133} $	&	1	&	WKP13	\\
24	&	No.37	& $	1.851	^{+0.848}_{-0.307}$ & $	1.176	^{+0.176}_{-0.301} $	&	0.1-1	&	WKP13	\\
25	&	No.38	& $	1.857	^{+0.310}_{-0.326}$ & $	1.041	^{+0.135}_{-0.196} $	&	1	&	WKP13	\\
26	&	No.39	& $	1.863	^{+0.316}_{-0.319}$ & $	1.041	^{+0.135}_{-0.196} $	&	1	&	WKP13	\\
27	&	No.40	& $	1.892	^{+0.309}_{-0.324}$ & $	1.176	^{+0.125}_{-0.176} $	&	1	&	WKP13	\\
28	&	No.42	& $	1.903	^{+0.325}_{-0.347}$ & $	0.602	^{+0.176}_{-0.301} $	&	1	&	WKP13	\\
29	&	No.43	& $	1.903	^{+0.330}_{-0.372}$ & $	0.699	^{+0.204}_{-0.398} $	&	2	&	WKP13	\\
30	&	No.44	& $	1.929	^{+0.232}_{-0.531}$ & $	0.919	^{+0.423}_{-0.074} $	&	0.1-1	&	WKP13	\\
31	&	No.46	& $	1.959	^{+0.306}_{-0.306}$ & $	0.978	^{+0.101}_{-0.133} $	&	0.1-1	&	WKP13	\\
32	&	No.47	& $	1.964	^{+0.313}_{-0.330}$ & $	0.845	^{+0.133}_{-0.192} $	&	1	&	WKP13	\\
33	&	No.48	& $	1.991	^{+0.312}_{-0.319}$ & $	1.176	^{+0.125}_{-0.176} $	&	1	&	WKP13	\\
34	&	No.49	& $	2.021	^{+0.076}_{-0.092}$ & $	0.996	^{+0.080}_{-0.098} $	&	1	&	WKP13	\\
35	&	No.50	& $	2.021	^{+0.313}_{-0.314}$ & $	0.699	^{+0.079}_{-0.097} $	&	1-3	&	WKP13	\\
36	&	No.52	& $	2.045	^{+0.311}_{-0.313}$ & $	0.978	^{+0.101}_{-0.133} $	&	1	&	WKP13	\\
37	&	MYSO 052124	& $	2.061	^{+0.126}_{-0.126}$ & $	1.150	^{+0.130}_{-0.130} $	&	1-2.5	&	2	\\
38	&	MYSO 045403	& $	2.146	^{+0.126}_{-0.126}$ & $	1.150	^{+0.130}_{-0.130} $	&	1-2.5	&	2	\\
39	&	No.56	& $	2.176	^{+0.308}_{-0.307}$ & $	1.270	^{+0.111}_{-0.156} $	&	2.5	&	WKP13	\\
40	&	No.57	& $	2.179	^{+0.326}_{-0.366}$ & $	0.699	^{+0.204}_{-0.398} $	&	1	&	WKP13	\\
41	&	No.59	& $	2.290	^{+0.400}_{-0.433}$ & $	1.090	^{+0.165}_{-0.245} $	&	1	&	WKP13	\\
42	&	MYSO 053244	& $	2.322	^{+0.126}_{-0.126}$ & $	1.150	^{+0.130}_{-0.130} $	&	1-2.5	&	2	\\
43	&	No.60	& $	2.352	^{+0.312}_{-0.319}$ & $	1.176	^{+0.125}_{-0.176} $	&	0-3	&	WKP13	\\
44	&	No.61	& $	2.378	^{+0.304}_{-0.303}$ & $	1.041	^{+0.135}_{-0.196} $	&	1	&	WKP13	\\
45	&	MYSO 053431	& $	2.455	^{+0.126}_{-0.126}$ & $	1.150	^{+0.130}_{-0.130} $	&	1-2.5	&	2	\\
46	&	No.63	& $	2.467	^{+0.310}_{-0.641}$ & $	1.176	^{+0.125}_{-0.176} $	&	1	&	WKP13	\\
47	&	MYSO 050941	& $	2.484	^{+0.126}_{-0.126}$ & $	1.200	^{+0.130}_{-0.130} $	&	1-2.5	&	2	\\
48	&	MYSO 051906	& $	2.633	^{+0.126}_{-0.126}$ & $	1.320	^{+0.130}_{-0.130} $	&	1-2.5	&	2	\\
49	&	No.66	& $	2.664	^{+0.326}_{-0.308}$ & $	1.079	^{+0.125}_{-0.176} $	&	<3	&	WKP13	\\
50	&	MYSO 053342	& $	2.703	^{+0.126}_{-0.126}$ & $	1.490	^{+0.130}_{-0.130} $	&	1-2.5	&	2	\\
51	&	No.69	& $	2.720	^{+0.306}_{-0.309}$ & $	1.041	^{+0.135}_{-0.196} $	&	4	&	WKP13	\\
52	&	No.70	& $	2.720	^{+0.307}_{-0.309}$ & $	1.439	^{+0.129}_{-0.138} $	&	3	&	WKP13	\\
53	&	No.71	& $	2.729	^{+0.309}_{-0.314}$ & $	1.207	^{+0.256}_{-0.362} $	&	2.5	&	WKP13	\\
54	&	No.72	& $	2.761	^{+0.303}_{-0.303}$ & $	1.540	^{+0.141}_{-0.160} $	&	2-3	&	WKP13	\\
55	&	No.73	& $	2.772	^{+0.311}_{-0.315}$ & $	1.320	^{+0.142}_{-0.174} $	&	2-3	&	WKP13	\\
56	&	No.76	& $	2.825	^{+0.302}_{-0.301}$ & $	1.439	^{+0.129}_{-0.138} $	&	2	&	WKP13	\\
57	&	No.77	& $	2.839	^{+0.306}_{-0.307}$ & $	1.170	^{+0.292}_{-0.537} $	&	0.5	&	WKP13	\\
58	&	No.81	& $	2.908	^{+0.305}_{-0.306}$ & $	1.470	^{+0.308}_{-0.169} $	&	3	&	WKP13	\\
59	&	No.85	& $	3.048	^{+0.304}_{-0.305}$ & $	1.744	^{+0.095}_{-0.111} $	&	2.3	&	WKP13	\\
60	&	No.86	& $	3.070	^{+0.306}_{-0.536}$ & $	1.978	^{+0.119}_{-0.165} $	&	1-3	&	WKP13	\\
61	&	No.87	& $	3.073	^{+0.484}_{-0.609}$ & $	1.041	^{+0.105}_{-0.196} $	&	1-3	&	WKP13	\\
62	&	No.88	& $	3.102	^{+0.301}_{-0.301}$ & $	1.744	^{+0.095}_{-0.111} $	&	1	&	WKP13	\\
63	&	No.89	& $	3.149	^{+0.327}_{-0.302}$ & $	1.477	^{+0.176}_{-0.301} $	&	<4	&	WKP13	\\
64	&	No.90	& $	3.151	^{+0.327}_{-0.311}$ & $	1.400	^{+0.144}_{-0.169} $	&	1-3	&	WKP13	\\
65	&	No.92	& $	3.158	^{+0.306}_{-0.307}$ & $	1.354	^{+0.151}_{-0.178} $	&	3-4	&	WKP13	\\
66	&	No.93	& $	3.162	^{+0.305}_{-0.311}$ & $	1.176	^{+0.125}_{-0.176} $	&	1-3	&	WKP13	\\
67	&	No.94	& $	3.226	^{+0.307}_{-0.308}$ & $	1.354	^{+0.151}_{-0.178} $	&	4	&	WKP13	\\
68	&	No.95	& $	3.232	^{+0.303}_{-0.301}$ & $	1.702	^{+0.097}_{-0.123} $	&	1-3	&	WKP13	\\
69	&	No.96	& $	3.239	^{+0.304}_{-0.305}$ & $	1.398	^{+0.170}_{-0.167} $	&	2	&	WKP13	\\
70	&	No.99	& $	3.327	^{+0.306}_{-0.308}$ & $	1.540	^{+0.141}_{-0.160} $	&	<1	&	WKP13	\\
71	&	No.100	& $	3.352	^{+0.303}_{-0.304}$ & $	1.601	^{+0.123}_{-0.154} $	&	<1	&	WKP13	\\
72	&	No.103	& $	3.364	^{+0.311}_{-0.306}$ & $	1.439	^{+0.129}_{-0.138} $	&	2	&	WKP13	\\
73	&	No.105	& $	3.378	^{+0.393}_{-0.975}$ & $	1.041	^{+0.105}_{-0.196} $	&	<4	&	WKP13	\\
74	&	No.106	& $	3.417	^{+0.305}_{-0.307}$ & $	1.439	^{+0.129}_{-0.138} $	&	2	&	WKP13	\\
75	&	No.107	& $	3.423	^{+0.304}_{-0.305}$ & $	1.595	^{+0.103}_{-0.104} $	&	2	&	WKP13	\\
76	&	No.108	& $	3.503	^{+0.300}_{-0.294}$ & $	1.354	^{+0.151}_{-0.178} $	&	4	&	WKP13	\\
77	&	No.110	& $	3.567	^{+0.307}_{-0.305}$ & $	2.000	^{+0.176}_{-0.097} $	&	2-4	&	WKP13	\\
78	&	No.111	& $	3.574	^{+0.320}_{-0.312}$ & $	1.655	^{+0.101}_{-0.137} $	&	3	&	WKP13	\\
79	&	No.112	& $	3.586	^{+0.304}_{-0.303}$ & $	1.933	^{0.146+}_{-0.127} $	&	1	&	WKP13	\\
80	&	No.114	& $	3.655	^{+0.309}_{-0.301}$ & $	1.654	^{+0.145}_{-0.207} $	&	<5	&	WKP13	\\
81	&	No.115	& $	3.656	^{+0.305}_{-0.306}$ & $	1.572	^{+0.110}_{-0.125} $	&	2	&	WKP13	\\
82	&	No.116	& $	3.656	^{+0.305}_{-0.307}$ & $	1.954	^{+0.125}_{-0.176} $	&	3-4	&	WKP13	\\
83	&	No.117	& $	3.662	^{+0.305}_{-0.304}$ & $	1.623	^{+0.296}_{-0.092} $	&	1	&	WKP13	\\
84	&	No.118	& $	3.670	^{+0.310}_{-0.307}$ & $	1.548	^{+0.133}_{-0.168} $	&	2.5	&	WKP13	\\
85	&	No.119	& $	3.673	^{+0.307}_{-0.308}$ & $	1.540	^{+0.141}_{-0.160} $	&	1.5	&	WKP13	\\
86	&	No.120	& $	3.697	^{+0.331}_{-0.317}$ & $	1.439	^{+0.129}_{-0.138} $	&	2-4	&	WKP13	\\
87	&	No.122	& $	3.726	^{+0.304}_{-0.303}$ & $	1.602	^{+0.122}_{-0.155} $	&	4.2	&	WKP13	\\
88	&	No.125	& $	3.774	^{+0.307}_{-0.309}$ & $	1.744	^{+0.095}_{-0.111} $	&	1.9	&	WKP13	\\
89	&	No.126	& $	3.830	^{+0.307}_{-0.305}$ & $	1.668	^{+0.218}_{-0.100} $	&	3	&	WKP13	\\
90	&	No.127	& $	3.836	^{+0.447}_{-0.313}$ & $	1.702	^{+0.097}_{-0.123} $	&	<5	&	WKP13	\\
91	&	No.128	& $	3.842	^{+0.302}_{-0.301}$ & $	2.079	^{+0.097}_{-0.079} $	&	1-3	&	WKP13	\\
92	&	No.130	& $	3.947	^{+0.305}_{-0.304}$ & $	2.083	^{+0.093}_{-0.195} $	&	1-3	&	WKP13	\\
93	&	No.131	& $	3.985	^{+0.302}_{-0.301}$ & $	1.828	^{+0.091}_{-0.104} $	&	1.5	&	WKP13	\\
94	&	No.132	& $	4.037	^{+0.306}_{-0.308}$ & $	1.756	^{+0.089}_{-0.093} $	&	2	&	WKP13	\\
95	&	No.133	& $	4.078	^{+0.301}_{-0.334}$ & $	1.818	^{+0.235}_{-0.055} $	&	1	&	WKP13	\\
96	&	No.135	& $	4.253	^{+0.310}_{-0.301}$ & $	1.999	^{+0.177}_{-0.221} $	&	1.7	&	WKP13	\\
97	&	No.136	& $	4.403	^{+0.303}_{-0.300}$ & $	1.790	^{+0.085}_{-0.083} $	&	1.3	&	WKP13	\\
98	&	No.137	& $	4.880	^{+0.309}_{-0.307}$ & $	1.964	^{+0.212}_{-0.138} $	&	2	&	WKP13	\\
99	&	No.138	& $	4.888   ^{+0.301}_{-0.308}$ & $ 2.045   ^{+0.131}_{-0.200} $	&	<4.4 &	WKP13	\\
100	&	No.139	& $	5.348   ^{+0.303}_{-0.303}$ & $ 2.098   ^{+0.078}_{-0.195} $	&	1-2.5 &	WKP13	\\
\hline
\end{longtable}
\tablebib{(1)~\citet{2011ApJ...727...64K}; (2) \citet{2017ApJ...834...94S}.}
\tablefoot{Here shows the cluster designation given by the reference source, empirical cluster masses ($M_{\mathrm{ecl}}$), maximal star masses ($m_{\rm max}$) within these clusters, cluster ages, and the references for the data in the order of star cluster mass for the compiled clusters as described in Section~\ref{sec: data} and \ref{sec: Mass uncertainties} and plotted in Fig.~\ref{fig: RSMmaxMecl} and \ref{fig:MmaxMeclRS2_opt}. }
}

\end{document}